\documentclass[amssymb,prd,superscriptaddress,aps,nofootinbib,twocolumn,showpacs]{revtex4-1}
\usepackage{graphicx, epsfig, amssymb} 
\usepackage{amsmath, amsfonts}
\usepackage{bm} 
\usepackage[breaklinks]{hyperref}
\usepackage{color}
\usepackage{enumerate}

\def\nn{\nonumber}

\def\be{\begin{equation}}
\def\ee{\end{equation}}
\def\beq{\begin{eqnarray}}
\def\eeq{\end{eqnarray}}
\def\lp{{\ell+1}}
\def\lm{{\ell-1}}
\def\lpp{{\ell+2}}
\def\lmm{{\ell-2}}

\def\th{\vartheta}

\def\cQ{{\cal Q}}

\begin{document}

\title{Tidal deformations of a spinning compact object}

\author{Paolo Pani}\email{paolo.pani@roma1.infn.it}
\affiliation{Dipartimento di Fisica, ``Sapienza'' Universit\`a di Roma \& Sezione INFN Roma1, P.A. Moro 5, 00185, Roma, Italy}
\affiliation{CENTRA, Departamento de F\'{\i}sica, Instituto Superior T\'ecnico, Universidade de Lisboa, Avenida~Rovisco Pais 1, 1049 Lisboa, Portugal.}

\author{Leonardo Gualtieri}\email{leonardo.gulatieri@roma1.infn.it}
\affiliation{Dipartimento di Fisica, ``Sapienza'' Universit\`a di Roma \& Sezione INFN Roma1, P.A. Moro 5, 00185, Roma, Italy}

\author{Andrea Maselli}\email{andrea.maselli@roma1.infn.it}
\affiliation{Center for Relativistic Astrophysics, School of Physics,
Georgia Institute of Technology, Atlanta, Georgia 30332, USA}

\author{Valeria Ferrari}\email{valeria.ferrari@roma1.infn.it}
\affiliation{Dipartimento di Fisica, ``Sapienza'' Universit\`a di Roma \& Sezione INFN Roma1, P.A. Moro 5, 00185, Roma, Italy}

\begin{abstract} 
The deformability of a compact object induced by a perturbing tidal field is encoded in the tidal Love numbers, which depend sensibly on the object's internal structure. These numbers are known only for static, spherically-symmetric objects. As a first step to compute the tidal Love numbers of a spinning compact star, here we extend powerful perturbative techniques to compute the exterior geometry of a spinning object distorted by an axisymmetric tidal field to second order in the angular momentum. The spin of the object introduces couplings between electric and magnetic deformations and new classes of induced Love numbers emerge. For example, a spinning object immersed in a quadrupolar, electric tidal field can acquire some induced mass, spin, quadrupole, octupole and hexadecapole moments to second order in the spin. The deformations are encoded in a set of inhomogeneous differential equations which, remarkably, can be solved analytically in vacuum. We discuss certain subtleties in defining the tidal Love numbers in general relativity, which are due to the difficulty in separating the tidal field from the linear response of the object in the solution, even in the static case. By extending the standard procedure to identify the linear response in the static case, we prove analytically that the Love numbers of a Kerr black hole remain zero to second order in the spin. As a by-product, we provide the explicit form for a slowly-rotating, tidally-deformed Kerr black hole to quadratic order in the spin, and discuss its geodesic and geometrical properties.
\end{abstract}

\pacs{
04.20.-q,  	
04.25.-g,	
04.70.Bw,	
04.30.-w.	
}

\maketitle

\section{Introduction}

\subsection{Context of this work}

The relativistic theory of the tidal deformations of a compact object and the dynamics of a tidally-distorted self-gravitating body are fascinating and challenging problems in general relativity, which have received considerable attention in recent years. In a binary system at large orbital separation, the tidal interaction is negligible and the two objects can be treated as point particles. However, as the orbit shrinks due to gravitational-wave emission, tidal interactions become increasingly important, and deform the multipolar structure of each object, their gravitational field, and the orbital motion up to the merger, where tidal deformations are dramatic and cannot longer be treated as small perturbations. Such tidal effects can leave a detectable imprint~\cite{Flanagan:2007ix,Hinderer:2007mb} in the gravitational waveform emitted by a neutron-star binary in the late stages of its orbital evolution, this system being the main target of current-generation gravitational-wave detectors~\cite{aLIGO,aVIRGO,KAGRA}.

The multipole moments of a compact object in a binary system are deformed by the tidal field produced by its companion. In Newtonian gravity, the constants of proportionality between the multipole moments of a mass distribution and the perturbing external tidal field in which the object is immersed are known as tidal Love numbers (see e.g.~\cite{Murraybook,PoissonWill}). 
These numbers depend sensibly on the object's internal structure and can thus provide a mean to understand the physics of neutron-star cores at ultranuclear density~\cite{Postnikov:2010yn,Lattimer:2004pg}. 
Motivated by the prospect of measuring the Love numbers through gravitational-wave detections~\cite{Flanagan:2007ix,Hinderer:2007mb,Hinderer:2009ca,Baiotti:2010xh,Baiotti:2011am,Vines:2011ud,Pannarale:2011pk,Lackey:2011vz,Damour:2012yf,Vines:2010ca,Lackey:2013axa,Favata:2013rwa,Yagi:2013baa}, and also by the need of improving gravitational-wave templates (e.g.~\cite{Bernuzzi:2012ci,Read:2013zra}, cf.~\cite{Buonanno:2014aza} for a review), in recent years a relativistic theory of tidal Love numbers has been developed with considerable success~\cite{Hinderer:2007mb,Binnington:2009bb,Damour:2009vw,Damour:2009va,Landry:2014jka,Guerlebeck:2015xpa}. The tidal Love numbers of static neutron stars have been computed for various realistic equations of state with great precision~\cite{Hinderer:2009ca,Postnikov:2010yn}, whereas Schwarzschild black holes (BHs) were shown to have precisely zero tidal Love numbers~\cite{Binnington:2009bb} and, therefore, the multipolar structure of a Schwarzschild BH is unaffected by a perturbing tidal field. Very recently, this result was extended beyond the perturbative level~\cite{Guerlebeck:2015xpa}.

These investigations have experienced a second burst of activity since the discovery of a set of relations among the moment of inertia, the electric tidal Love number, and the spin-quadrupole moment of a neutron star, which were remarkably found to be almost independent of the equation of state~\cite{Yagi:2013bca,Yagi:2013awa}. These relations have been studied and extended in various scenarios~\cite{Maselli:2013mva,Doneva:2013rha,Haskell,Maselli:2013rza,Chakrabarti:2013tca,Sham:2013cya,Pani:2014jra,Yagi:2014qua,Sham:2014kea,Chan:2014tva,Yagi:2015hda}
and allow us to compute two elements of the triad once a single one is measured. For example, a single measurement of the tidal Love number of a neutron star would allow us --~provided independent measurements of the mass and spin are available~-- to compute the moment of inertia and the spin-quadrupole moment. 
Further, universal relations have also been discovered, relating the mass quadrupole moment, the current octupole moment, and higher-order multipole moments~\cite{Yagi:2013sva,Pappas:2013naa,Stein:2014wpa,Yagi:2014bxa,Yagi:2015upa}.

A further motivation to study tidal interactions in gravitational systems comes from the relativistic dynamics of an extreme-mass ratio inspiral, i.e. a stellar-size compact object being captured by a supermassive BH, which is one of the main targets of future space-based detectors~\cite{AmaroSeoane:2012je}. In the late stages of the inspiral before the plunge, the small object can perform hundreds of thousands of cycles around the supermassive BH, deforming the geometry of the latter by an amount proportional to the mass ratio~\cite{Hughes:2001jr,Price:2001un,Yunes:2009ef,Chatziioannou:2012gq}. This translates into changes to the orbital motion and to the gravitational-wave phase which are comparable to self-force~\cite{Barack:2009ux} effects.

So far, the relativistic tidal Love numbers have been considered only for nonspinning objects. 
This is clearly a strong limitation, because compact objects in binary systems are expected to possess a nonvanishing angular momentum, and also because the coupling between the spin and the tidal field can produce novel effects that have been neglected in previous works.

In this paper, we start a long-term effort to compute the tidal Love numbers of a spinning compact object and to study the tidal deformability in the presence of spin. 
Some complementary aspects of this complex problem have been investigated in the past. The geometry of a tidally-deformed, spinning BH was obtained in Ref.~\cite{Yunes:2005ve} by integrating the Teukolsky equations, whereas Ref.~\cite{O'Sullivan:2014cba} studied the intrinsic geometry of a spinning BH distorted by a small compact companion in the extreme-mass ratio limit. The latter two papers considered arbitrary values of the BH spin parameter.
Very recently, Poisson computed the geometry of a slowly-spinning BH distorted by a quadrupolar electric and magnetic tidal field to first order in the BH spin~\cite{Poisson:2014gka}. He decomposed the
tidal perturbations in terms of irreducible potentials and extracted the tidal moments through a matching to the post-Newtonian metric of a binary system. Although perturbative in the spin, the work of Ref.~\cite{Poisson:2014gka} provides physical insight on the spin-tidal coupling and presents the deformed solution in a simple and elegant way.

\subsection{Executive summary}

Here we summarize our main results and compare our work with previous analyses. Given the complexity and richness of the
problem, the purpose of the present paper is manifold. First, we wish to develop a framework to compute the tidal Love
numbers of spinning compact stars and we present here the first necessary step in that direction. This requires going
beyond the analyses of Refs.~\cite{Yunes:2005ve,O'Sullivan:2014cba,Poisson:2014gka}, which only considered tidal
deformations of spinning BHs. Similarly to Ref.~\cite{Poisson:2014gka}, our framework is that of relativistic
perturbations of a spinning geometry in the slow-rotation approximation [developed in Sec.~\ref{sec:slowrot} and in
  Appendix~\ref{app:Kojima}]. Our results are valid to second order in the object's angular momentum. A perturbative
expansion in the spin is a reasonable assumption, first because neutron stars in binary systems are expected to rotate
slowly, and also because the quadrupole moment (which describes the most relevant tidal deformations) is quadratic in the angular momentum. 

We find the general solution of Einstein's equations
describing --~in a closed analytical form~-- the exterior of a slowly-spinning object deformed by a tidal field generated by distant sources. Both the
inner object and the distant sources are stationary and axisymmetric, and Einstein's equations are linearized in the
perturbations induced by the tidal field.  Since the general solution is a linear combination of independent solutions,
it depends on a number of arbitrary constants. The values of these constants should be fixed by matching with
the solution describing the interior of the compact star, or by imposing regularity conditions at the horizon if the central object is a BH.

The structure of the general solution before fixing the constants allows us to discriminate the terms of the spacetime
metric describing the tidal field from those describing the linear response of the compact object to the latter.  This
discrimination is crucial to properly define the mass and current multipole moments~\cite{Geroch:1970cd,Hansen:1974zz}
($M_\ell$ and $S_\ell$, respectively) of the central object in the buffer
zone~\cite{Thorne:1980ru,Thorne:1984mz,Hinderer:2007mb,Binnington:2009bb}.  Once the multipole moments are determined,
the Love number can be obtained as the coefficients of proportionality between the multipole moments and the tidal
field, i.e. we define the quantities
\begin{equation}
 \lambda_\ell^{(M)}\equiv\frac{\partial M_\ell}{\partial{\cal E}_0}\,,\qquad \lambda_\ell^{(S)}\equiv\frac{\partial S_\ell}{\partial{\cal E}_0}\,, \label{defLove}
\end{equation}
where ${\cal E}_0$ is the amplitude of the tidal field. Because we include only linear corrections in the tidal field,
the quantities above are independent of ${\cal E}_0$.

As we discuss in Sec.~\ref{subsec:separation}, discriminating between the tidal field and the linear response of the
compact object is a delicate issue, especially in the rotating case. Indeed, there is always the freedom to incorporate
part of the response solution into the external tidal field, thus shifting the values of the multipole moments and then
of the Love numbers (see e.g. \cite{Thorne:1984mz,Fang:2005qq,Damour:2009va,Kol:2011vg}). To fix this ambiguity, here we follow a prescription, which we believe
is reasonable and it is also consistent with previous literature on the subject
\cite{Hinderer:2007mb,Binnington:2009bb,Damour:2009vw,Poisson}. However, this prescription is
more well founded when the inner object is a BH than when it is a star. A more complete investigation of this delicate
issue is left for a future publication; in particular, it is important to understand whether the Love numbers obtained
with this choice correspond to actual measurable quantities, such for instance those appearing in post-Newtonian
waveforms.

For our general solution, we extract the Geroch-Hansen multipole moments~\cite{Geroch:1970cd,Hansen:1974zz} and compute the
formal expressions for the tidally-induced changes of the first relevant mass and current moments from Eq.~\eqref{defLove}.  As discussed in
detail in Sec.~\ref{sec:slowrot}, the spin of the object introduces couplings between electric and magnetic deformations
and new classes of induced Love numbers emerge [cf. Eqs.~\eqref{Love0}--\eqref{Love4}]. In a future publication, the
results presented here will be used to explicitly compute the tidal Love numbers~\eqref{Love0}--\eqref{Love4} of a
spinning neutron star to second order in the spin.

When the object is a BH, the metric simplifies considerably. In this case, we impose the
boundary conditions by requiring that the metric is regular at the horizon. The explicit solution that we present in
Sec.~\ref{sec:BH} and Appendix~\ref{app:BHsol} extends that derived in Ref.~\cite{Poisson:2014gka} to include
second-order effects in the spin but --~at variance with the solution of Ref.~\cite{Poisson:2014gka}~-- is limited to
the axisymmetric case and to the case in which the magnetic quadrupolar component of the tidal tensor is zero. As a
by-product of our analysis, we compute a novel family of tidal Love numbers of the Kerr BH. Using the separation of tidal and response solutions discussed above [cf. Sec.~\ref{subsec:separation}], we find that the multipole moments of a tidally-distorted
Kerr geometry are unaffected by the tidal source to second order in the spin and therefore the corresponding Love numbers~\eqref{defLove} vanish [see discussion in Sec.~\ref{sec:LoveBH}]. This result extends the work of Ref.~\cite{Binnington:2009bb}, which found that the tidal
Love numbers of a Schwarzschild BH are zero~\cite{BirthOfIdea}\footnote{Very recently, the no-hair properties of tidally-distorted Schwarzschild BHs have been proved also beyond the perturbative regime~\cite{Guerlebeck:2015xpa}.}. Working independently from us, Landry and Poisson have recently obtained a complementary result~\cite{Poisson}, namely they found that the Love numbers of a Kerr BH are zero to first order in the spin but, differently from ours, their results also include nonaxisymmetric tidal perturbations and a quadrupolar magnetic tidal field. 

Finally, our analytical BH solution can be used to compute
easily the near-horizon deviations from a Kerr geometry induced by the external tidal field. In Sec.~\ref{sec:BH}, we
provide analytical expressions for the epicyclic frequencies, curvature invariants and other geometrical quantities
which can be relevant to investigate the dynamics of a binary system containing a spinning BH. We conclude in
Sec.~\ref{sec:conclusions} by discussing various interesting extensions of our work.

\subsection{Working assumptions}\label{sec:setup}
The relativistic dynamics of a spinning object immersed in a tidal field is a challenging problem and we make a number of working assumptions to treat it. For the reader's convenience, we list these assumptions below:
\begin{enumerate}[(i)]
 \item The central object is slowly spinning, with dimensionless spin parameter $\chi\equiv J/M^2\ll1$, where $M$ and $J$ are the object's mass and angular momentum, respectively. Our analysis is valid to ${\cal O}(\chi^2)$. Neglecting rotation and tidal deformations, the central is object is spherically symmetric. 
 \item The tidal field varies slowly in time, so that time derivatives in Einstein's equations are small compared to spatial derivatives. In practice, we consider stationary tidal perturbations.
 \item The object's spin axis coincides with the axis of symmetry of the tidal field, which is assumed to be axisymmetric. This prevents the tidal field and the spin axis from precessing~\cite{Thorne:1997kt}, thus leaving the geometry stationary. By adopting this assumption, here we mostly focus on the axisymmetric case, $m=0$, where $m$ is the azimuthal number of the tidal perturbations. Our method might be extended to the nonaxisymmetric case (describing, for instance, the complete tidal field generated by an orbiting companion), although precession introduces time dependence in the problem. We note that, because the background is axisymmetric, perturbations with different values of $m$ are decoupled from each other. Therefore, our approach can describe the axisymmetric tidal response of the object, regardeless of the symmetries of the external tidal field.
 \item The sources of the tidal field are localized at large distance $r_0$. Our vacuum solution is valid in the region $R<r\ll r_0$, where $R$ is the radius of the object.
 \item The tidal field is weak: our results are valid to linear order in the amplitude of the tidal field. Assuming the tidal field is generated by a ring of mass $m_c$ [as to satisfy the hypothesis of axisymmetry of point (iii) above], our results are valid for $m_c M^2/r_0^3\ll1$.
 \item For simplicity, we assume that the tidal field is mostly electric and quadrupolar and consider $\ell=2$ polar tidal perturbations in the static case, where $\ell$ is the standard harmonic index. This assumption is not crucial and can be easily lifted by including different $\ell$ multipoles and a magnetic component.
\end{enumerate}
Through this work, we use $G=c=1$ units.

\section{Framework: slow-rotation expansion}\label{sec:slowrot}

The linearized dynamics of a spinning perturbed object in general relativity is a challenging problem because of mode mixing in the perturbation equations. However, a perturbative expansion in the angular velocity of the object can render the problem tractable. 
Our framework is that of linear perturbations of a slowly-rotating star, that has been initiated in Refs.~\cite{ChandraFerrari91,Kojima:1992ie,1993ApJ...414..247K,1993PThPh..90..977K,Ferrari:2007rc} and recently extended and put on firmer basis in the context of BH perturbations~\cite{Pani:2012vp,Pani:2012bp} (see~\cite{Pani:2013pma} for a review). 

The basic idea is that slowly-rotating geometries are ``close enough'' to spherical symmetry that an approximate separation of the perturbation equations in radial and angular parts becomes possible. The perturbation functions are expanded in spherical harmonics and they reduce to a system of differential equations where various couplings between different multipolar indices $\ell$ and between perturbations with different parity are introduced. 
The slow-rotation approximation imposes some selection rules on couplings between different multipoles and only a certain number of higher multipoles contributes to a given order in $\epsilon_a=\Omega/\Omega_K\ll1$, where $\Omega$ is the uniform angular velocity of the star and $\Omega_K=\sqrt{M/R^3}$ is the mass-shedding frequency\footnote{When considering BHs, the expansion parameter is naturally $\epsilon_a=\chi$. Hereafter we use $\epsilon_a$ as a bookkeeping parameter for the expansion in the angular momentum.}. This makes the method well suited to investigate complicated systems of coupled equations, which can be finally integrated by standard methods~\cite{Pani:2013pma}. 

We consider a stationary, external tidal gravitational field which is characterized in terms of the Weyl's tensor $C_{ijkl}$ evaluated in the local rest frame, ${\cal E}_{ij}=C_{i0j0}$ and ${\cal B}_{ij}=\frac{1}{2}\epsilon_{ipq}C^{pq}_{j0}$ for the electric and magnetic tidal quadrupole moments, respectively. Because of our working assumptions, it suffices to consider only \emph{stationary and axisymmetric} perturbations of the spinning background, although our technique can be generalized to dynamical perturbations of any spinning compact objects to arbitrary order in the spin.

\subsection{Spinning background}
Following Hartle and Thorne~\cite{Hartle:1967he,Hartle:1968si}, the
most general stationary and axisymmetric\footnote{We also require the spacetime to be symmetric with respect to the equatorial plane, and invariant under the ``circularity condition'', $t\to-t$ and $\varphi\to-\varphi$, which implies $g_{t\vartheta}=g_{t\varphi}=g_{r\vartheta}=g_{r\varphi}=0$~\cite{Chandra}. Note that, while the circularity condition follows from Einstein and Maxwell equations in electrovacuum, it might not hold true for other matter fields.} metric $g_{\mu\nu}$ to ${\cal O}(\epsilon_a^2)$ in rotation can be written as
\begin{eqnarray}
 ds^2&&=-e^\nu\left[1+2\epsilon_a^2\left(j_0+j_2 P_2\right)\right]dt^2\nn\\
 &&+\frac{1+2\epsilon_a^2(m_0+m_2P_2)/(r-2{\cal M})}{1-2{\cal M}/r}dr^2\nn\\
 &&+r^2\left[1+2\epsilon_a^2(v_2-j_2)P_2\right]\left[d\vartheta^2+\sin^2\vartheta(d\varphi-\epsilon_a\omega dt)^2\right]\,,\nn\\ \label{metric}
\end{eqnarray}
where $P_2=P_2(\cos\vartheta)=(3\cos^2\vartheta-1)/2$ is a Legendre
polynomial. The radial functions $\nu$ and ${\cal M}$ are of zeroth order in
rotation, $\omega$
is of first order, and $j_0$, $j_2$,
$m_0$, $m_2$, $v_2$ are of second order. We also introduce the functions $e^{-\lambda}=1-2{\cal M}/r$ and $\bar\omega=\Omega-\omega$  at zeroth and first order in the rotation, respectively.
%
By plugging this decomposition into the
gravitational equations $R_{\mu\nu}=0$ and by solving the
equations order by order in $\epsilon_a$, we obtain a system of ordinary differential equations (ODEs) for the rotating background~\cite{Hartle:1967he,Hartle:1968si}.
To second order in the spin\footnote{This expansion has been recently extended to ${\cal O}(\epsilon_a^4)$~\cite{Yagi:2014bxa}.}, the vacuum background metric is given in Appendix~\ref{app:background}. This solution describes the exterior of a spinning object, with mass $M$ in the static case and dimensionless spin parameter $\chi\ll1$. As discussed below, the constants $\Omega$, $\delta m$ and $\delta q$ which appear in the solution are related to the angular velocity of the central object and to ${\cal O}(\chi^2)$- corrections to the mass and and to the quadrupole moment, respectively. 

The geometry given in Appendix~\ref{app:background} describes a slowly-rotating Kerr BH as a particular case. 
In this case, regularity of the metric at the Schwarzschild horizon $r=2M$ imposes $\delta q=0$, whereas $\delta m$ can be set to zero without loss of generality through a mass rescaling. Our coordinates differ from the more standard Boyer-Lindquist ones. In the metric~\eqref{metric} the horizon's location is $r_+=2M\left[1-\epsilon_a^2\chi^2/8\right]$, whereas the ergoregion is located at $r_{\rm ergo}=2M\left[1-\epsilon_a^2\chi^2\cos(2\vartheta)/8\right]$. Although the coordinates are singular at $r=2M$, all curvature invariants are regular outside the horizon.

\subsection{Perturbations}

Slowly rotating and oscillating compact objects can be studied as perturbations of the axisymmetric, stationary solutions discussed above.
For completeness, here we discuss the case of nonaxisymmetric perturbations with azimuthal number $m$ and harmonic index $\ell\geq|m|$. Later, we shall restrict to $m=0$.
Scalar, vector and tensor field equations in the background
metric~\eqref{metric} can be linearized in the field perturbations.
Any perturbation function $\delta X$ can be expanded in a complete basis of spherical harmonics; schematically, we have
\begin{equation}
\delta X_{\mu_1\dots}(t,r,\vartheta,\varphi)=
\delta X^{(i)}_{\ell m}(r)\hat{\cal Y}_{\mu_1\dots}^{\ell m\,(i)}(\th,\varphi)\,,
\label{expa}
\end{equation}
where we have imposed that the perturbations do not depend on $t$ explicitly and $\hat{\cal Y}_{\mu_1\dots}^{\ell m\,(i)}$ is a basis of scalar,
vector or tensor harmonics, depending on the tensorial nature of the
perturbation $\delta X$. As in the spherically symmetric case, the perturbation variables $\delta X^{(i)}_{\ell m}$ can
be classified as ``polar'' or ``axial'' depending on their behavior
under parity transformations.

The linear response of the system is fully characterized by a coupled system of ODEs in the perturbation functions $\delta X^{(i)}_{\ell  m}$.  
In the case of a spherically symmetric background,
perturbations with different values of $(\ell,\,m)$, as well as
perturbations with opposite parity, are decoupled. In a rotating,
axially symmetric background, perturbations with different values of
$m$ are still decoupled but perturbations with different values of
$\ell$ are not.

To second order, the perturbation equations read schematically (see Ref.~\cite{Pani:2013pma} for a pedagogical derivation)
\begin{eqnarray}
0&=&{\cal A}_{\ell}+\epsilon_a m \bar{\cal A}_{{\ell}}+\epsilon_a^2 \hat{{\cal A}}_\ell\nn\\
&+&\epsilon_a ({\cal Q}_{{\ell}}\tilde{\cal P}_{\ell-1}+{\cal Q}_{\ell+1}\tilde{\cal P}_{\ell+1})\nn\\
&+&\epsilon_a^2 \left[\cQ_\lm \cQ_\ell \breve{{\cal A}}_\lmm + \cQ_\lpp \cQ_\lp 
\breve{{\cal A}}_\lpp \right]+{\cal O}(\epsilon_a^3)\,,\label{eq_axial}\\
0&=&{\cal P}_{\ell}+\epsilon_a m \bar{\cal P}_{{\ell}}+\epsilon_a^2 \hat{{\cal P}}_\ell\nn\\
&+&\epsilon_a ({\cal Q}_{{\ell}}\tilde{\cal A}_{\ell-1}+{\cal Q}_{\ell+1}\tilde{\cal A}_{\ell+1})\nn\\
&+&\epsilon_a^2 \left[\cQ_\lm \cQ_\ell \breve{{\cal P}}_\lmm + \cQ_\lpp \cQ_\lp 
\breve{{\cal P}}_\lpp \right]+{\cal O}(\epsilon_a^3)\,,\label{eq_polar}
\end{eqnarray}
where we have defined 
\begin{equation}
 {\cal Q}_\ell=\sqrt{\frac{\ell^2-m^2}{4\ell^2-1}}\,,\label{Qpm}
\end{equation}
and ${\cal A}_{\ell}$, $\bar {\cal A}_{\ell}$, $\tilde {\cal A}_{\ell}$,
$\hat {\cal A}_{\ell}$, $\breve {\cal A}_{\ell}$ are \emph{linear}
combinations of the axial perturbations with
multipolar index $\ell$; similarly, ${\cal P}_{\ell}$, $\bar {\cal
  P}_{\ell}$, $\tilde {\cal P}_{\ell}$, $\hat {\cal P}_{\ell}$,
$\breve {\cal P}_{\ell}$ are linear combinations of the polar
perturbations with index $\ell$. 

The structure of Eqs.~\eqref{eq_axial}--\eqref{eq_polar} is interesting.
In the limit of slow rotation a Laporte-like
``selection rule''~\cite{ChandraFerrari91} imposes perturbations with a given parity and index $\ell$ to couple only to: (i)
perturbations with \emph{opposite} parity and index $\ell\pm1$ at ${\cal O}(\epsilon_a)$; (ii) perturbations with \emph{same} parity and
\emph{same} index $\ell$ up to ${\cal O}(\epsilon_a^2)$; (iii)
perturbations with \emph{same} parity and index $\ell\pm2$ at
${\cal O}(\epsilon_a^2)$. 
Furthermore, from Eq.~\eqref{Qpm} it follows that ${\cal Q}_{\pm m}=0$, and
therefore if $|m|=\ell$ the coupling of perturbations with index
$\ell$ to perturbations with indices $\ell-1$ and $\ell-2$ is
suppressed. This general property is known as ``propensity
rule''~\cite{ChandraFerrari91} in atomic theory, and states that transitions $\ell\to\ell+1$
are strongly favored over transitions $\ell\to\ell-1$. Note that the slow-rotation technique is well known in quantum mechanics and the coefficients ${\cal Q}_{\ell}$ are in fact related to the usual Clebsch-Gordan coefficients~\cite{Pani:2013pma}.

\subsection{Axial-led and polar-led perturbations}
Due to the coupling between different multipolar indices, Eqs.~\eqref{eq_axial}--\eqref{eq_polar} form an infinite system of coupled ODEs and the spectrum of their solutions is extremely rich. 
However, in special configurations the perturbation equations can be greatly simplified, as we now show. 

First, we expand the axial and polar perturbation functions (schematically denoted as $a_{\ell m}$ and $p_{\ell m}$, respectively) that appear in Eqs.~\eqref{eq_axial} and \eqref{eq_polar}:
\begin{eqnarray}
a_{\ell m}&=&a^{(0)}_{\ell m}+\epsilon_a\,a^{(1)}_{\ell m}+\epsilon_a^2a^{(2)}_{\ell m}+{\cal O}(\epsilon_a^3)\nn\\
p_{\ell m}&=&p^{(0)}_{\ell m}+\epsilon_a\,p^{(1)}_{\ell m}+\epsilon_a^2p^{(2)}_{\ell m}+{\cal O}(\epsilon_a^3)\,.
\end{eqnarray}
The terms $\breve{{\cal A}}_{\ell\pm2}$ and $\breve{{\cal P}}_{\ell\pm2}$ in Eqs.~\eqref{eq_axial}--\eqref{eq_polar} are multiplied by factors $\epsilon_a^2$, so they
only depend on the zeroth-order perturbation functions, $a^{(0)}_{\ell\pm2
  m}$, $p^{(0)}_{\ell\pm2 m}$. The terms $\tilde{{\cal A}}_{\ell\pm1}$ and
$\tilde{{\cal P}}_{\ell\pm1}$ are multiplied by factors $\epsilon_a$,
so they only depend on zeroth- and first-order perturbation functions
$a^{(0)}_{\ell\pm1 m}$, $p^{(0)}_{\ell\pm1 m}$, $a^{(1)}_{\ell\pm1 m}$,
$p^{(1)}_{\ell\pm1 m}$.

Since in the nonrotating limit axial and polar perturbations are
decoupled, a possible consistent set of solutions of the system
\eqref{eq_axial}--\eqref{eq_polar} has $a^{(0)}_{{L}\pm2 m}\equiv0$ and $p^{(0)}_{{L}\pm1 m}\equiv0$, where $\ell=L$ is a specific value of the harmonic index. This ansatz leads to the ``axial-led''~\cite{Lockitch:1998nq} subset of Eqs.~\eqref{eq_axial}--\eqref{eq_polar}:
\begin{equation}\label{axial_led}
 \left\{ \begin{array}{l}
          {\cal A}_{{L}}+\epsilon_a m \bar{\cal A}_{{{L}}}+\epsilon_a^2 \hat{{\cal  A}}_{L}+\epsilon_a ({\cal Q}_{{{L}}}\tilde{\cal P}_{{L}-1}+
{\cal Q}_{{L}+1}\tilde{\cal P}_{{L}+1})=0\\
{\cal P}_{{L}+1}+\epsilon_a m \bar{\cal P}_{{{L}+1}}+
\epsilon_a {\cal Q}_{{{L}+1}}\tilde{\cal A}_{{L}}=0\\
{\cal P}_{{L}-1}+\epsilon_a m \bar{\cal P}_{{{L}-1}}+
\epsilon_a {\cal Q}_{{{L}}}\tilde{\cal A}_{{L}}=0\\
{\cal A}_{L+2}+\epsilon_a {\cal Q}_{{{L+2}}}\tilde{\cal P}_{{L}+1}+\epsilon_a^2 {\cal Q}_{L+1} {\cal Q}_{L+2} \breve{\cal A}_L=0\\
{\cal A}_{L-2}+\epsilon_a {\cal Q}_{{{L-1}}}\tilde{\cal P}_{{L}-1}+\epsilon_a^2 {\cal Q}_{L} {\cal Q}_{L-1} \breve{\cal A}_L=0
         \end{array}\right.\,, 
\end{equation}
where the first equation is solved to second order in the spin, the second and the third equations do not contain zeroth-order quantities in the spin, and the last two equations do not contain zeroth- and first-order terms in the spin, i.e. $a_{L\pm2}={\cal O}(\epsilon_a^2)$. The truncation above is consistent because in the axial equations for $\ell=L$ the polar source terms with $\ell=L\pm1$ appear multiplied by a factor $\epsilon_a$, so terms $p_{L\pm1 m}^{(2)}$ would be of higher order in the axial equations.

Similarly, another consistent set of solutions of the same system has
$p^{(0)}_{{L}\pm2 m}\equiv0$ and $a^{(0)}_{{L}\pm1 m}\equiv0$. The corresponding ``polar-led'' system reads 
\begin{equation}\label{polar_led}
 \left\{ \begin{array}{l}
          {\cal P}_{{L}}+\epsilon_a m \bar{\cal P}_{{{L}}}+\epsilon_a^2 \hat{{\cal P}}_{L}+
\epsilon_a ({\cal Q}_{{{L}}}\tilde{\cal A}_{{L}-1}+{\cal Q}_{{L}+1}
\tilde{\cal A}_{{L}+1})=0\\
{\cal A}_{{L}+1}+\epsilon_a m \bar{\cal A}_{{{L}+1}}
+\epsilon_a {\cal Q}_{{{L}+1}}\tilde{\cal P}_{{L}}=0\\
{\cal A}_{{L}-1}+\epsilon_a m \bar{\cal A}_{{{L}-1}}+
\epsilon_a {\cal Q}_{{{L}}}\tilde{\cal P}_{{L}}=0\\
{\cal P}_{L+2}+\epsilon_a {\cal Q}_{{{L+2}}}\tilde{\cal A}_{{L}+1}+\epsilon_a^2 {\cal Q}_{L+1} {\cal Q}_{L+2} \breve{\cal P}_L=0\\
{\cal P}_{L-2}+\epsilon_a {\cal Q}_{{{L-1}}}\tilde{\cal A}_{{L}-1}+\epsilon_a^2 {\cal Q}_{L} {\cal Q}_{L-1} \breve{\cal P}_L=0
         \end{array}\right.\,.
\end{equation}
Interestingly, within this perturbative scheme a notion of ``conserved quantum number'' ${L}$ is still meaningful: even though, for any given ${L}$, rotation couples terms with opposite parity and different multipolar index, the subsystems~\eqref{axial_led} and \eqref{polar_led} are closed, i.e. they contain a \emph{finite} number of equations which fully describe the dynamics to second order in the spin.

The main assumption that leads to Eq.~\eqref{axial_led} [resp. Eq.~\eqref{polar_led}]  is that only axial (resp. polar) perturbations with harmonic index ${L}$ are activated at zeroth order in the rotation. In terms of an external tidal field, we are assuming that such field is a pure $\ell=L$ magnetic (resp. electric) state at zeroth order in the rotation. This assumption would not hold if the zeroth-order tidal field is a mixture between different $\ell$ states. In such case one has to deal with the full system~\eqref{eq_axial}--\eqref{eq_polar}, which is much more involved. However, working with the system~\eqref{polar_led} should provide a reliable approximation, because the electric quadrupolar ($\ell=2$ ) contribution to the external tidal field is the dominant one.

The explicit form of the axial-led and polar-led systems~\eqref{axial_led} and ~\eqref{polar_led} for a spinning stationary and axisymmetric object is derived in Appendix~\ref{app:Kojima} and is available in a {\scshape Mathematica}\textsuperscript{\textregistered} notebook provided in the Supplemental Material.

\section{Tidal deformations to second-order in the spin} \label{sec:tidal}
Let us now focus on axisymmetric ($m=0$) stationary perturbations, which are a particular case of those discussed above and in Appendix~\ref{app:Kojima}. 
The axial-led (resp. polar-led) system~\eqref{axial_led} [resp.~\eqref{polar_led}] describes the second-order spin corrections to a nonrotating object immersed in an $\ell=L$ magnetic (resp. electric) tidal field, which we now compute. 
For clarity, we focus on the dominant $L=2$ polar-led perturbations, although the same procedure can be applied to other values of $L$ and to axial-led perturbations. As discussed above, $L=2$ polar perturbations in the static case couple to $L=1$ and $L=3$ axial perturbations to first order in the spin, and to $L=0$, $L=2$ and $L=4$ polar perturbations to second order in the spin. Within this perturbative scheme, the perturbation equations at each order are naturally written as inhomogeneous ODEs, where the homogeneous part depends on differential operators defined in the nonrotating case, whereas the sources depend on the couplings between different perturbations. For this reason, it is useful to present the equations in the static case for generic values of $L$.

Unless otherwise written, henceforth we reabsorb the bookkeeping parameter $\epsilon_a$ in $\chi$.

\subsection{Tidal perturbations at zeroth order}
In the static case, axial and polar perturbations as well as perturbations with different harmonic indices are decoupled from each other. To zeroth order in the spin, the axial and polar sectors of stationary perturbations with $\ell=L\geq2$ reduce to two single decoupled homogeneous ODEs~\cite{Hinderer:2007mb,Binnington:2009bb}:
\begin{eqnarray}
 {\cal D}_{P,L}[{H_0^{(L)}}]&&\equiv {H_0^{(L)}}''+\frac{2 (r-M)}{r (r-2 M)} {H_0^{(L)}}'\nn\\
 &&-\frac{L (L+1) r (r-2 M)+4 M^2}{r^2 (r-2 M)^2}{H_0^{(L)}}=0\,, \label{eqH0zeroth}\\
 {\cal D}_{A,L}[{h_0^{(L)}}]&&\equiv {h_0^{(L)}}''+\left[\frac{4 M-r L(L+1)}{r^3(r-2M)}\right]{h_0^{(L)}}=0\,, \nn\\\label{eqh0zeroth}
\end{eqnarray}
where we have defined the differential operators ${\cal D}_{P,L}$ and ${\cal D}_{A,L}$.
The metric function $K^{(L)}$ is algebraically related to $H_0^{(L)}$ and its derivatives, whereas the functions $H_1^{(L)}$ and $h_1^{(L)}$ vanish at zeroth order in the spin and therefore the polar perturbations are also static. Focusing for simplicity on the $L=2$ polar sector, the explicit solution reads~\cite{Hinderer:2007mb,Binnington:2009bb}
\begin{eqnarray}
 H_0^{(2)}&=&\frac{3 \alpha  r (r-2 M)}{M^2}+\frac{\gamma }{2}  \left[\frac{3 r (r-2 M)}{M^2} \log \left(1-\frac{2 M}{r}\right)\right.\nn \\
 &&\left.-\frac{2 M}{r}+r \left(\frac{6}{M}-\frac{1}{r-2M}\right)-5\right]\,,\label{polL2spin0}\\
 H_2^{(2)}&=& H_0^{(2)}\,,\\
 K^{(2)}&=&3 \alpha  \left(\frac{r^2}{M^2}-2\right)+\gamma  \left[\left(\frac{3 r^2}{2 M^2}-3\right) \log \left(1-\frac{2 M}{r}\right)\right.\nn\\
 &&\left.+\frac{3 r}{M}-\frac{2 M}{r}+3\right] \label{K0inf}\,,
 \end{eqnarray}
where $\alpha$ and $\gamma$ are two integration constants, so that the solution above is a linear combination of two independent solutions.

The solution proportional to $\alpha$ diverges at large distances and can be identified with the external tidal field~\cite{Thorne:1980ru,Hinderer:2007mb,Binnington:2009bb}, whereas the solution proportional to $\gamma$ naturally represents the object's linear response to the applied tidal perturbation (see however Sec.~\ref{subsec:separation} for a more detailed discussion). As we shall discuss later, the constant $\gamma$ is associated with the tidally-induced quadrupole moment of the object in the static case, whereas $\alpha$ is proportional to the axisymmetric component of the electric quadrupolar tidal field (roughly speaking, $\alpha\sim m_c M^2/r_0^3$ for a ring of mass $m_c$ at orbital distance $r_0$). 
The ratio $\gamma/\alpha$ is proportional to the $L=2$ electric Love number~\cite{Hinderer:2007mb,Binnington:2009bb},
\begin{equation}
 k_{\rm el}^{(2)}= -\frac{4 \gamma  M^5}{15 \alpha R^5}\,, \label{LoveNumber}
\end{equation}
where $R$ is the radius of the object.
The constants $\alpha$ and $\gamma$ are proportional to each other, their ratio being determined by matching the exterior solution above to the regular solution describing the perturbed object's interior. Thus, for an object of given mass and composition, the Love number $k_{\rm el}^{(2)}$ is uniquely determined. In the static case, the dependence of the Love number on the stellar equation of state has been discussed in detail in Refs.~\cite{Flanagan:2007ix,Hinderer:2007mb,Binnington:2009bb,Hinderer:2009ca,Postnikov:2010yn,Yagi:2013bca,Yagi:2013awa}.

In case the central object is a BH, regularity of the geometry at the Schwarzschild horizon $r_+=2M$ imposes $\gamma=0$, as can be directly seen from Eq.~\eqref{polL2spin0} or by computing some curvature invariant at the horizon. Therefore, the Love number~\eqref{LoveNumber} of a tidally-deformed Schwarzschild BH is precisely zero. 
This is a general result, which is valid for any $L$ and for electric and magnetic tidal perturbations~\cite{Binnington:2009bb}. In other words, the multipolar structure of a static BH is not deformed by a perturbing tidal field (very recently, this result has been extended to arbitrary values of a static tidal field~\cite{Guerlebeck:2015xpa}). 

%
\subsection{First-order corrections}

The zeroth-order solution~\eqref{polL2spin0}--\eqref{K0inf} sources the axial perturbations with $L=1$ and $L=3$ through the second and third equations in the system~\eqref{polar_led}, yielding
\begin{eqnarray}
 {\cal D}_{A,1}[{h_0^{(1)}}]&\equiv&{h_0^{(1)}}''-\frac{2}{r^2}{h_0^{(1)}}= S_{A}^{(1)} \,, \label{eqAXIALl1}\\ 
 {\cal D}_{A,2}[{h_0^{(3)}}]&\equiv&{h_0^{(3)}}''-\frac{4 (3r-M)}{r^2 (r-2 M)}{h_0^{(3)}}= S_{A}^{(3)}\,, \label{eqAXIALl3}
\end{eqnarray}
where ${\cal D}_{A,i}$ are differential operators\footnote{Note that, while ${\cal D}_{A,2}$ is obtained from the operator ${\cal D}_{A,L}$ for $L=2$, the operator ${\cal D}_{A,1}$ is different because $L=1$ perturbations satisfy a different set of equations, as discussed in Appendix~\ref{app:Kojima}.} and $S_{A}^{(L)}$ are source terms that are given in Appendix~\ref{app:sources}. The axial metric functions $h_1^{(3)}$ and $h_1^{(1)}$ vanish identically.
As expected, the sources are proportional to the coupling between the background gyromagnetic term, $g_{t\varphi}$, and the zeroth-order function $H_0^{(2)}$. It is easy to verify that the first-order corrections to the polar perturbations are vanishing, so Eq.~\eqref{eqH0zeroth} and the two equations above fully characterize the polar-led $L=2$ system to first order in the spin.
The explicit solution of the equations above is given in Appendix~\ref{app:solutions}. This solution depends on four new integration constants, $\alpha_{1,3}$ and $\gamma_{1,3}$, which arise from the homogeneous problem associated with Eqs.~\eqref{eqAXIALl1} and~\eqref{eqAXIALl3}, and that are discussed in Sec.~\ref{subsec:solution}.


\subsection{Second-order corrections}

With the zeroth-order and the first-order solutions at hand, from the first equation in the system~\eqref{polar_led} we can compute the second-order correction to the metric coefficient $H_0^{(2)}(r)$, which we denote by $\delta H_0^{(2)}(r)$ to distinguish it from the zeroth order quantity.
This correction satisfies the following inhomogeneous ODE
\begin{eqnarray}
 {\cal D}_{P,2}[&\delta H_0^{(2)}&]\equiv {\delta H_0^{(2)}}''+  \frac{2 (r-M)}{r(r-2 M)} {\delta H_0^{(2)}}' \nn\\
 &-& \frac{2 \left(2 M^2+3 r^2-6 M r\right)}{r^2 (r-2 M)^2} \delta H_0^{(2)} = S_P^{(2)}\,,  \label{eqBHP2}
\end{eqnarray}
where the source $S_P^{(2)}$ is also given in Appendix~\ref{app:sources}. Note that $\delta H_1^{(2)}=0$ [i.e. polar perturbations remain static also to ${\cal O}(\chi^2)$], whereas $\delta H_2^{(2)}$ and $\delta K^{(2)}$ are algebraically related to $\delta H_0^{(2)}$ and its derivatives.
Finally, to fully characterize the second-order corrections, one needs to compute the last two equations in~\eqref{polar_led}, which define the second-order terms in the induced $L=0$ and $L=4$ polar sectors.
The $L=4$ system reduces to the second-order ODE
\begin{eqnarray}
 {\cal D}_{P,4}[&\delta H_0^{(4)}&]\equiv {\delta H_0^{(4)}}''+  \frac{2 (r-M)}{r(r-2 M)} {\delta H_0^{(4)}}' \nn\\
 &-& \frac{4 \left(M^2+5 r^2-10 M r\right)}{r^2 (r-2 M)^2} \delta H_0^{(4)} = S_P^{(4)}\,, \label{eqBHP4}
\end{eqnarray}
where the source $S_P^{(4)}$ is given in Appendix~\ref{app:sources}. Also in this case the other $L=4$ polar components follow algebraically from $\delta H_0^{(4)}$ and its derivatives.

On the other hand, the $L=0$ polar system satisfies a different set of equations (cf. Appendix~\ref{app:Kojima}), which can be reduced to the following first-order system:
\begin{eqnarray}
 {\delta H_0^{(0)}}'+\frac{\delta H_2^{(0)}}{r-2M}= S_P^{(0,0)} \,, \label{eqP0a}\\
 {\delta H_2^{(0)}}'+\frac{\delta H_2^{(0)}}{r-2M}= S_P^{(0,2)} \,, \label{eqP0b}
\end{eqnarray}
and the sources $S_P^{(0,0)}$ and $S_P^{(0,2)}$ are given in Appendix~\ref{app:sources}.

Remarkably, all the equations above can be solved analytically. Schematically, the nonvanishing metric coefficients to quadratic order in the spin read (reinstating the bookkeeping parameter $\epsilon_a$ only in these equations)
\begin{eqnarray}
 g_{tt} &=& -e^\nu\left[1+2\epsilon_a^2\left(j_0+j_2 P_2-\frac{r^2 e^{-\nu}}{2}(\Omega-\bar\omega)^2\right) \right.\nn\\
 &&\left. +\epsilon_a^2 \delta H_0^{(0)} Y^{00}+\left(H_0^{(2)}+\epsilon_a^2\delta H_0^{(2)}\right) Y^{20}\right.\nn\\
 &&\left. + \epsilon_a^2 \delta H_0^{(4)} Y^{40} \right] \,, \label{metricF1}\\
 g_{t\varphi} &=& -\epsilon_a r^2 (\Omega-\bar\omega) \sin^2\vartheta\nn\\
 &&+\epsilon_a \sin\vartheta\left(h_0^{(1)} Y^{10}_{,\vartheta}+h_0^{(3)} Y^{30}_{,\vartheta}\right)\,,\\
 g_{rr} &=& \left[1-\frac{2{\cal M}}{r}\right]^{-1}\times\left[1+2\epsilon_a^2\frac{m_0+m_2P_2}{r-2{\cal M}}\right.\nn\\
 &&\left. +\epsilon_a^2 \delta H_2^{(0)} Y^{00}+\left(H_2^{(2)}+\epsilon_a^2\delta H_2^{(2)}\right) Y^{20}\right.\nn\\
 &&\left. + \epsilon_a^2 \delta H_2^{(4)} Y^{40} \right] \,,\\
 g_{\vartheta\vartheta} &=& r^2\left[1+2\epsilon_a^2(v_2-j_2)P_2\right.\nn\\
 &&\left. +\epsilon_a^2 \delta K^{(0)} Y^{00}+\left(K^{(2)}+\epsilon_a^2\delta K^{(2)}\right) Y^{20}\right.\nn\\
 &&\left. + \epsilon_a^2 \delta K^{(4)} Y^{40} \right] \,,\\
 g_{\varphi\varphi} &=& \sin^2\vartheta g_{\vartheta\vartheta}\,, \label{metricF5}
\end{eqnarray}
where we recall that $Y^{\ell0}=Y^{\ell0}(\vartheta)$ are the scalar spherical harmonics with $m=0$ and $P_2\equiv 2\sqrt{\pi/5}Y^{20}$ is a Legendre polynomial.
The radial functions $\nu$, ${\cal M}$, $\bar\omega$, $j_0$, $j_2$, $m_0$, $m_2$, $v_2$ are given in Appendix~\ref{app:background}; the radial functions $h_0^{(1)}$ and $h_0^{(2)}$ are given in Appendix~\ref{app:solutions}; the radial functions $H_0^{(2)}$, $H_2^{(2)}$ and $K^{(2)}$ are given in Eqs.~\eqref{polL2spin0}--\eqref{K0inf}; whereas the radial functions $\delta H_0^{(0)}$, $\delta H_0^{(2)}$, $\delta H_0^{(4)}$, $\delta H_2^{(0)}$, $\delta H_2^{(2)}$, $\delta H_2^{(4)}$, $\delta K^{(0)}$, $\delta K^{(2)}$, $\delta K^{(4)}$ are cumbersome and, to avoid typographical errors and help comparison, their full solution is provided in an online notebook in the Supplemental Material. Note that the only nonvanishing off-diagonal term of the metric is $g_{t\varphi}$ and it only contains the background gyromagnetic term and the axial perturbations with $L=1$ and $L=3$.

\subsection{Description of the solution}\label{subsec:solution}
Equations~\eqref{metricF1}--\eqref{metricF5} fully describe the exterior metric of a tidally-deformed spinning object to second order in the spin and in the region $R<r\ll r_0$.
Because the procedure to obtain such solution is considerably involved, as a nontrivial consistency check we have verified that the explicit solution satisfies Einstein's equations in vacuum, $R_{\mu\nu}=0$, to quadratic order in the spin and to linear order in the tidal perturbations.

The background Hartle-Thorne solution depends on the parameters $M$, $\chi$, $\Omega$, $\delta m$ and $\delta q$; the ${\cal O}(\chi^0)$ tidal solution depends on the constants $\alpha$ and $\gamma$; the ${\cal O}(\chi)$ tidal solution depends on the constants $\alpha_{1,3}$ and $\gamma_{1,3}$; finally, the ${\cal O}(\chi^2)$ tidal solution depends on the constants $\alpha_{0,2,4}$ and $\gamma_{0,2,4}$. The full solution depends on 17 free parameters, whose physical meaning is summarized in Table~\ref{tab:metric} and will be discussed in more detail below. 
Note that each pair of constants $\alpha_\ell$ and $\gamma_\ell$ arises from the homogeneous problem associated with the inhomogeneous equations for the corresponding multipole $\ell$ presented above.

As a representative example, we present here the explicit form of the function $\delta H_0^{(0)}$, which is the most compact among the second-order perturbations:
\begin{widetext}
 \begin{eqnarray}
 \delta H_0^{(0)} &=& \alpha_0\chi^2+\frac{ \gamma_0 \chi ^2}{2-y}+ \frac{8\alpha_1 \chi^2}{\sqrt{3} (y-2)}+ \frac{2\gamma_1 \chi^2}{\sqrt{3} (y-2) y^3}+\alpha \chi^2\left[\frac{4 (6+y (2y-1) (1+y (6y-1)))}{\sqrt{5} (y-2) y^4}\right.\nn\\
 &&\left.+\delta q\left(\frac{3 \left(-4+y (-4+(y-2) y (3 (y-1) y-14))+3 (y-2)^3 y^2 (2+y) \coth^{-1}[1-y]\right)}{\sqrt{5} (y-2) y}\right)\right]\nn\\
 &&+\gamma\chi^2\left[\frac{36-y (26+3 y (2+(y-1) y (15y-11)))+\frac{3}{2} y (24+y (-12+y (12+y (-32+(41-15 y) y)))) \log\left[\frac{y-2}{y}\right]}{3
   \sqrt{5} (y-2) y^5} \right.\nn\\
 &&\left.+
\frac{\delta q}{2 \sqrt{5} (y-2)^2 y^2}  
 \left(24+2 y \left(-20+y \left(256+9 y \left(-14+(y-2)^2 y\right)\right)\right)\right.\right.\nn\\
 &&\left.\left.+6 (y-2) y (-4+y (-6+(y-2) y (3 (y-1) y-14)))
   \log\left[\frac{y-2}{y}\right]+\frac{9}{2} (y-2)^4 y^3 (2+y) \log\left[\frac{y-2}{y}\right]^2\right)\right]\,, \label{H0}
\end{eqnarray}
\end{widetext}
where $y=r/M$.
The parameters $\alpha_0$ and $\gamma_0$ arise from the homogeneous problem associated with Eqs.~\eqref{eqP0a}--\eqref{eqP0b}.
The constant $\alpha_0$ only appears in $g_{tt}$ and, as we discuss below, can be eliminated through a time
rescaling. The constants $\alpha_1$ and $\gamma_1$ are related to the source terms proportional to the axial term with
$L=1$, $h_0^{(1)}$, which sources $H_0^{(0)}$ in Eqs.~\eqref{eqP0a}--\eqref{eqP0b} through a coupling with the spin of
the object, whereas the constants $\alpha$ and $\gamma$ are related to the polar terms with $L=2$ at zeroth order in the spin (namely $H_0^{(2)}$, $H_2^{(2)}$ and $K^{(2)}$) which couples to $\delta H_0^{(0)}$
at second order in the spin. Clearly, the structure of the solution reflects the selection rules discussed in
Sec.~\ref{sec:slowrot}.

More generically, the tidal corrections of the metric can be schematically written as a linear combination of independent solutions in the following form:
\begin{equation}
 \delta g_{\mu\nu} = \alpha \delta g_{\mu\nu}^{(\alpha)}+\gamma \delta g_{\mu\nu}^{(\gamma)}+\sum_{\ell=0}^4\left[\alpha_\ell \delta g_{\mu\nu}^{(\alpha_\ell)}+\gamma_\ell \delta g_{\mu\nu}^{(\gamma_\ell)}\right]\,,\label{decomposition}
\end{equation}
where we have factored out the dependence on $\alpha$, $\gamma$, $\alpha_\ell$ and $\gamma_\ell$. The first two
functions $\delta g_{\mu\nu}^{(\alpha)}$ and $\delta g_{\mu\nu}^{(\gamma)}$ contain terms of zeroth order in
the spin and also terms of first and second order in the spin which arise as the particular solutions of the
inhomogeneous equations; the functions $\delta g_{\mu\nu}^{(\alpha_\ell)}$, $\delta g_{\mu\nu}^{(\gamma_\ell)}$ with odd $\ell$ contain ${\cal O}(\chi)$ terms arising from the solutions of the
homogeneous problem and also ${\cal O}(\chi^2)$ terms arising from the particular solutions of the inhomogeneous
equations; finally, the functions $\delta g_{\mu\nu}^{(\alpha_\ell)}$, $\delta g_{\mu\nu}^{(\gamma_\ell)}$ with even $\ell$ contain only ${\cal O}(\chi^2)$ terms.

\begin{table}[t]
\caption{List of the free parameters appearing in the tidally-deformed metric of a spinning vacuum geometry for polar-led $L=2$ perturbations. The subscript in $\alpha_\ell$ and $\gamma_\ell$ refers to the multipole that is related to the specific constant. The precise relation between the multipole moments and the constants $\gamma$ and $\gamma_\ell$ is given in Eqs.~\eqref{moment0}--\eqref{moment4}.}
\begin{tabular}{cc|l}
 \hline\hline
${\cal O}(\chi,\alpha)$  & & \\
\hline
\hline
$(0,0)$ & $M$		&	mass \\
$(1,0)$ & $\chi$	&	spin \\
$(1,0)$ & $\Omega$	&	angular velocity of the object \\
$(2,0)$ & $\delta m$	&	spin-induced mass shift \\
$(2,0)$ & $\delta q$	&	spin-induced quadrupole-moment shift\\
 \hline
$(0,1)$ & $\alpha$	&	external electric quadrupolar tidal field \\
$(0,1)$ & $\gamma$	&	static response to the external tidal field \\
 \hline
$(1,1)$ &  $\alpha_1$	&	external magnetic dipolar tidal field \\
$(1,1)$ &  $\gamma_1$	&	tidally-induced spin shift \\
  \hline
$(2,1)$ &  $\alpha_0$	&	constant shift of $g_{tt}$ at infinity \\
$(2,1)$ &  $\gamma_0$	&	tidally-induced mass shift \\
$(2,1)$ &  $\alpha_2$	&	external electric quadrupolar tidal field  \\
$(2,1)$ &  $\gamma_2$	&	tidally-induced quadrupole-moment shift \\
$(2,1)$ &  $\alpha_3$	&	external magnetic octupolar tidal field  \\
$(2,1)$ &  $\gamma_3$	&	tidally-induced octupole-moment shift  \\
$(2,1)$ &  $\alpha_4$	&	external electric $L=4$ tidal field \\
$(2,1)$ &  $\gamma_4$	&	tidally-induced hexadecapole-moment shift \\
\hline\hline
\end{tabular}
\label{tab:metric}
\end{table}
\subsubsection{Separating the tidal part from the linear response}\label{subsec:separation}

Crucially, the only parts of the metric which diverge in the far-field limit are $\delta g_{\mu\nu}^{(\alpha)}$ and $\delta g_{\mu\nu}^{(\alpha_\ell)}$, whereas the terms $\delta g_{\mu\nu}^{(\gamma)}$ and $\delta g_{\mu\nu}^{(\gamma_\ell)}$ yield an asymptotically-flat solution. However, note that one cannot simply identify the terms proportional to $\alpha$ and $\alpha_\ell$ with those associated with the external tidal field, and those proportional to $\gamma$ and $\gamma_\ell$  with the linear response of the central objects. Such definition suffers from an ambiguity because a trivial shift of the integration constants
\begin{equation}
 \gamma= \gamma'-\alpha \hat\gamma\,,\qquad \gamma_\ell=\gamma_\ell'-\alpha \hat\gamma_\ell\,, \label{shift}
\end{equation}
would transform Eq.~\eqref{decomposition} to the equivalent form\footnote{As discussed below, a quadrupole-led electric tidal field at infinity imposes $\alpha_\ell=0$. For clarity, we have already used this condition in Eq.~\eqref{decomposition2} and postpone its proof to Sec.~\ref{sec:boundary}.}
\begin{eqnarray}
 \delta g_{\mu\nu} &=& \alpha \left[\delta g_{\mu\nu}^{(\alpha)}-\hat\gamma \delta g_{\mu\nu}^{(\gamma)}-\sum_{\ell=0}^4\hat\gamma_\ell \delta g_{\mu\nu}^{(\gamma_\ell)}\right]\nn\\
 &&+\gamma' \delta g_{\mu\nu}^{(\gamma)}+\sum_{\ell=0}^4 \gamma_\ell'\delta g_{\mu\nu}^{(\gamma_\ell)}\,.\label{decomposition2}
\end{eqnarray}
While the new solution proportional to $\alpha$ is still divergent at large distances --~and therefore $\alpha$ can still be identified with the amplitude of the tidal field~-- the coefficients of the subleading solutions $\delta g_{\mu\nu}^{(\gamma)}$ and $\delta g_{\mu\nu}^{(\gamma_\ell)}$ have been shifted by an amount proportional to $\alpha$. In Sec.~\ref{sec:MM} below we show that such coefficients are related to the multipole moments of the central object (but see discussion in Sec.~\ref{sec:notemoments}). It is therefore crucial to find a unique prescription to characterize the linear response of the system.

In the nonrotating case, this ambiguity was mentioned in Refs.~\cite{Fang:2005qq,Damour:2009va,Kol:2011vg}. A rigorous way to identify the two solutions uniquely is to perform an analytical continuation in the number of spacetime dimensions $d$~\cite{Kol:2011vg} or in the multipolar index $\ell$~\cite{Chakrabarti:2013lua}\footnote{We are indebted to Jan Steinhoff for suggesting Refs.~\cite{Kol:2011vg,Chakrabarti:2013lua} to us.} and to recognize that the general solution for the gravitational potential at large distances schematically reads~\cite{Kol:2011vg,Chakrabarti:2013lua}
\begin{equation}
 \delta g_{tt} \sim  \alpha r^{\ell}\left(1+\dots\right)+ \gamma r^{-\ell-d+3}\left(1+\dots\right)\,,
\end{equation}
where the dots represent a series in $M/r$ which can also contain logarithmic terms and $\alpha$ and $\gamma$ are integration constants. Comparison with the Newtonian potential generated by an $\ell$-pole distribution allows us to identify the first and the second solution as those describing the tidal field and the linear response of the system, respectively.
By treating $d$ or $\ell$ as \emph{real} parameters, the two solutions above can always be distinguished without mixing of the possible common powers in the series expansion. Such procedure gives a precise proof of the identification done in Refs.~\cite{Hinderer:2007mb,Binnington:2009bb} for the nonrotating case. Indeed, it shows that the $\alpha$- and $\gamma$-solutions appearing in Eq.~\eqref{polL2spin0} precisely describe the tidal field and the linear response of the object, as anticipated.

Unfortunately, performing a similar analytical continuation in the rotating case  is impractical because the solution of Eq.~\eqref{polL2spin0} for generic $L$ is written in terms of Legendre functions, which appear in the sources of Eqs.~\eqref{eqAXIALl1}--\eqref{eqAXIALl3}. While, for any specific value of $L$, the latter equations admit a solution in closed form, the same does not seem true for generic $L$. Furthermore, the problem gets only more involved to second order in the spin. 
For this reason, to separate the tidal part from the linear response in the general solution~\eqref{decomposition} we have adopted a different procedure. We fix the shifts in Eq.~\eqref{shift} such that the new growing solution after the shift,
\begin{equation}
 \delta g_{\mu\nu}^{\rm growing}=\alpha \left[\delta g_{\mu\nu}^{(\alpha)}-\hat\gamma \delta g_{\mu\nu}^{(\gamma)}-\sum_{\ell=0}^4\hat\gamma_\ell \delta g_{\mu\nu}^{(\gamma_\ell)}\right]\,,
\end{equation}
only contains a \emph{finite} numbers of terms in a large-distance expansion. 
As it turns out, when the central object is a BH this prescription selects a \emph{unique} solution for the shifts~\eqref{shift}, at least up to quadratic order in the spin. We claim that such solution represents the physical solution of the tidal field, whereas the remaining part represents the physical response of the BH to the tidal field. 

In the case in which the central object is a compact star, this truncation only occurs up to first order in the spin. Indeed, to quadratic order there appear infinite terms proportional to $\alpha\chi^2 \delta q$ in the far-field expansion of $\delta g_{\mu\nu}^{\rm growing}$, which cannot be canceled by the shifts~\eqref{shift}. This reflects a certain arbitrariness in the definition of the Love numbers of a spinning star, which we will investigate in the future. In this case, we fix the shifts~\eqref{shift} such that the far-field expansion of the growing solution contains a finite number of terms, modulo those proportional to $\alpha\chi^2 \delta q$, which arise from the particular solution of the inhomogeneous problem at second order in the spin.

Although the one just described is admittedly not a rigorous prescription, it is nonetheless supported by the following
observations: (i) it allows us to identify correctly the two pieces of Eq.~\eqref{polL2spin0} in the nonrotating case, in
which the unique solution is known. Indeed, it is clear from Eq.~\eqref{polL2spin0} that $\hat\gamma=0$ is the only
possibility. (ii) To first order in the spin, it agrees with the recent results by Landry and Poisson~\cite{Poisson}
that were obtained using an independent prescription\footnote{We are indebted to Phil Landry and Eric Poisson for
  sharing their results with us which have helped in revising the argument above.}. In particular, our prescription
automatically implies that the tidal solution is regular in the exterior spacetime, also when the central object is a
BH. This is due to the fact that possible terms in the form $\sim\log(r-2M)$, $\sim1/(r-2M)$ that might appear in
$g_{\mu\nu}^{(\alpha)}$ are precisely removed using the prescription above.
Finally (iii), our procedure incorporates the idea that we ``superimpose'' a well-behaved tidal field on a central
object. The solution corresponding to the tidal field, regardless of whether the central object is a BH or a star,
should not have a pole at $r=2M$. We only allow poles arising from the coupling with irregular background terms, which
are those proportional to $\alpha\chi^2 \delta q$ discussed above.

In the rest of this paper, we will identify the tidal solution and the linear-response solution through the procedure just described and postpone a more rigorous analysis for future work. For simplicity, we rename the rescaled constants as $\gamma'\to \gamma$ and $\gamma_\ell'\to \gamma_\ell$. 
Note that the physical meaning of the constants listed in Table~\ref{tab:metric} refers to the solution after this rescaling.

Finally, we observe that although the tidal solution extracted through this procedure contains only a finite number of terms in the series expansion in the BH case, it can nonetheless contain terms with the same powers of $M/r$ as the linear-response solution. This is a peculiarity of the spinning solution and does not occur in the static case. For example, $\delta g_{\mu\nu}^{\rm growing}$ contains up to $M/r$ terms (which also appear in $\delta g_{\mu\nu}^{(\gamma_1)}$) to first order in the spin, and up to $(M/r)^5$ terms (which also appear in  $\delta g_{\mu\nu}^{(\gamma_4)}$) to second order in the spin.

\subsubsection{Boundary conditions at infinity} \label{sec:boundary}

The integration constants $\alpha_\ell$ that appear in the vacuum solution~\eqref{decomposition} have to be fixed by some
physical requirement on the nature of the tidal field. Because we assume an $(L=2)$-leading external tidal field, we fix
the constants $\alpha_\ell$ by imposing that the other leading $L\neq2$-contributions vanish. 

Let us start by the first-order corrections in the spin. The large-distance behavior of the gyromagnetic term reads
\begin{eqnarray}
 g_{t\varphi} &\to& \alpha_3  \chi M \left[\frac{1}{56}\frac{r^4}{M^4}-\frac{5}{84}\frac{r^3}{M^3}+\frac{1}{21}\frac{r^2}{M^2}\right] Y^{30}_{,\vartheta}\sin\vartheta\nn\\
 &&+\alpha_1 \chi \frac{r^2}{M} Y^{10}_{,\vartheta}\sin\vartheta + {\cal O}\left(\frac{r}{M}\right)\,.
\end{eqnarray}
The leading-order terms proportional to $Y^{30}_{,\vartheta}$ and $Y^{10}_{,\vartheta}$ would respectively correspond to some spurious $L=3$ and $L=1$ magnetic tidal perturbations~\cite{Binnington:2009bb} and we eliminate them by fixing $\alpha_3=0=\alpha_1$. Note that the constants $\gamma_1$ and $\gamma_3$ do not appear in the leading-order behavior at large distance and are therefore not associated to possible components of the tidal field.

Likewise, to second order in the spin, the dominant large-distance behavior of the $g_{tt}$ metric component reads
\begin{equation}
 g_{tt}\to \alpha_4 \chi^2 Y^{40}(\vartheta) \frac{r^4}{M^4}+ {\cal O}\left(\frac{r^3}{M^3}\right)\,.\nn
\end{equation}
This leading term is related to a possible $L=4$ component of the external electric tidal field, which we eliminate by fixing $\alpha_4=0$. Again, the constant $\gamma_4$ does not affect this leading-order behavior and it is indeed related only to nondivergent terms.

In a similar way, we fix $\alpha_2$ by requiring that the leading-order behavior of the component $g_{tt}\to r^2$ at large distances is not affected by the spin. This has to be the case because the leading-order behavior is related to the components of the tidal field, which does not depend on the properties of the central object, as discussed below.  
Finally, the constant $\alpha_0$ appearing in Eq.~\eqref{H0} can be eliminated through a time rescaling $t\to (1+\eta \chi^2)t$, where $\eta$ is a constant to be fixed. To second order in the spin, this rescaling only affects the coefficient $g_{tt}$ and not the gyromagnetic term $g_{t\varphi}$, because the latter is of the order $\chi$, so corrections would be cubic in $\chi$. Without loss of generality, we use this gauge freedom to fix 
\begin{equation}
 2\sqrt{\pi}\eta =-\alpha_0+\left(6\sqrt{5}\alpha-\frac{39}{\sqrt{5}}\gamma\right)\delta q\,,
\end{equation}
which completely cancels the term $\alpha_0$ in the metric and also simplifies the angular dependence of some subleading term in the $g_{tt}$ coefficient at large distances. 
Henceforth, we will enforce the time rescaling just described and that $\alpha_\ell=0$ ($\ell=1,..,4$). Thus, the free parameters of the tidal perturbations reduce to $\alpha$, $\gamma$ and $\gamma_\ell$ ($\ell=0,..,4$).

To check the asymptotic behavior of the solution, let us identify the tidal field in terms of the metric components. First, we consider the large-distance expansion of our solution after fixing the constants $\alpha_\ell$, 
\begin{eqnarray}
 1+g_{tt}&\to&  3 \alpha Y^{20}(\vartheta)\left[ \frac{r^2}{ M^2}-4 \left(1+{\delta m}  \chi ^2\right)\frac{r}{M} \right.\nn\\
    &&\left.+4   \left(1+2 {\delta m}  \chi ^2\right) \right]+\dots\,, \nn\\
 g_{t\varphi} &\to& -3\sqrt{\frac{5}{\pi}}\alpha \chi \sin^2\vartheta\left[r+\frac{3+5\cos(2\vartheta)}{4}\right] +\dots\,,\nn\\
 g_{rr}-1 &\to&  3 \alpha Y^{20}(\vartheta)\frac{r^2}{ M^2}+\dots\,,\nn\\
 \frac{g_{\vartheta\vartheta}}{r^2}-1&\to&  3 \alpha Y^{20}(\vartheta)\left[\frac{r^2}{ M^2}-8\sqrt{\pi}\left(1+2 {\delta m} \chi ^2\right)\right]+\dots\,,\nn
\end{eqnarray}
together with the exact relation $g_{\varphi\varphi} = \sin^2\vartheta g_{\vartheta\vartheta}$. In the expansion above, we neglected terms of the order $M/r$ or higher, which depend on $\alpha$, $\gamma$ and $\gamma_\ell$. Note that the time rescaling discussed above has been used both to eliminate $\alpha_0$ and to make the subleading, ${\cal O}(r^0)$, tidal term of $g_{tt}$ proportional to $Y^{20}(\vartheta)$. With a different gauge choice, the angular dependence would not factor out.

From the large-distance expansion above we can identify the tidal field. In a suitable gauge, the large-distance behavior of a vacuum spacetime distorted by a quadrupolar tidal field reads~\cite{Thorne:1984mz,Thorne:1997kt,Hinderer:2007mb,Binnington:2009bb}
\begin{eqnarray}
 g_{tt} &\to& -1-{\cal E}_{ij}x^i x^j+\dots\,, \label{tidalmetric1}\\
 g_{ti} &\to& -\frac{2}{3}\epsilon_{ijk} {{\cal B}^j}_l x^k x^l +\dots\,, \label{tidalmetric2}\\
 g_{ij} &\to&  \delta_{ij}\left[1-{\cal E}_{ij}x^i x^j\right]+\dots\,,\label{tidalmetric3}
\end{eqnarray}
where $\epsilon_{ijk}$ is the Levi-Civita symbol, ${\cal E}_{ij}$ and ${\cal B}_{ij}$ are the electric and magnetic tidal quadrupole moments, respectively, and $x^i$ are Cartesian coordinates. We can now transform to spherical coordinates, $x^i=r n^i$, with $n^i=(\sin\vartheta\cos\varphi,\sin\vartheta\sin\varphi,\cos\vartheta)$, and decompose the tidal tensor as~\cite{Hinderer:2007mb}
\begin{equation}
 {\cal E}_{ij}=\sum_{m=-2}^2 {\cal E}_m {\cal Y}_{ij}^{2m}(\vartheta,\varphi)\,,\label{decompE}
\end{equation}
where ${\cal Y}_{ij}^{2m}(\vartheta,\varphi)$ are symmetric traceless tensors~\cite{Thorne:1980ru} related to the usual spherical harmonics by $Y^{2m}(\vartheta,\varphi)={\cal Y}_{ij}^{2m}(\vartheta,\varphi) n^i n^j$. By plugging this decomposition into Eqs.~\eqref{tidalmetric1}--\eqref{tidalmetric3}, transforming the metric to spherical coordinates, and comparing with the leading-order asymptotic behavior of our solution, it is straightforward to identify
\begin{equation}
 {\cal E}_0=-\frac{3\alpha}{M^2}\,, \qquad  {\cal E}_{m\neq0}=0\,,\qquad {\cal B}_{ij}=0\,.\label{E0}
\end{equation}
The $g_{t\varphi}$ coefficient in the large-distance expansion above contains only subleading terms which are not related to a magnetic quadrupole tidal moment, but are due to the frame dragging effect. The fact that the only nonvanishing component of the tidal field is ${\cal E}_0$ is a consistency check of our solution, since we imposed a pure electric and axisymmetric quadrupolar tidal field. 
For a tidal source of mass $m_c$ at a distance $r_0$, ${\cal E}_0\sim m_c /r_0^3$~\cite{Taylor:2008xy,Poisson:2014gka} and Eq.~\eqref{E0} gives $\alpha\sim m_c M^2/r_0^3$, as previously anticipated.

\subsection{Multipole moments}\label{sec:MM}
%
In order to extract the multipole moments~\cite{Thorne:1980ru,PoissonWill} from our solution, we have to remove the tidal fields, as
in the nonrotating case \cite{Thorne:1984mz,Thorne:1997kt,Hinderer:2007mb,Binnington:2009bb}. This is done by identifying the solutions corresponding to the tidal field and to the linear response of the system as discussed in Sec.~\ref{subsec:separation} and by setting
to zero the constants corresponding to the unique growing solution thus defined, i.e., by fixing $\alpha=\alpha_\ell=0$. As discussed in Sec.~\ref{subsec:separation}, the remaining solution is asymptotically flat. In this way, we determine the spacetime's response to the tidal field\footnote{We remark that one cannot simply read off the multipole moments from the components of the full solution, following e.g. the prescription of~\cite{Thorne:1980ru}. Indeed, this prescription requires that the metric is expressed in asymptotically mass-centered Cartesian coordinates and, most importantly, that the spacetime is asymptotically flat. None of these conditions is satisfied by the full metric~\eqref{decomposition}.}.

We follow Ryan's approach to compute the Geroch-Hansen multipole moments~\cite{Geroch:1970cd,Hansen:1974zz} in a
gauge-invariant way through the geodesic properties of an axisymmetric, asymptotically-flat
spacetime~\cite{Ryan:1995wh,Ryan:1997hg,Pappas:2012nt}.
Ryan found that a low-velocity expansion of the energy change per logarithmic interval of the orbital frequency is completely determined by the multipole moments of a Ricci-flat solution. Such quantity is defined as
\begin{equation}
 \Delta E\equiv -\Omega \frac{\partial E_p}{\partial \Omega}\,,
\end{equation}
where $\Omega$ is the angular velocity of an equatorial circular geodesic with specific energy $E_p$. The large-distance expansion of $\Delta E$ reads~\cite{Ryan:1995wh,Yagi:2014bxa}
 \begin{eqnarray}
 \Delta E &&= \frac{v^2}{3} - \frac{v^4}{2}  + \frac{20}{9} \frac{S_1}{M_0^2}v^5 - \left( \frac{27}{8} - \frac{M_2}{M_0^3}  \right) v^6 + \frac{28}{3} \frac{S_1}{M_0^2}v^7 \nn\\
 &&- \left( \frac{225}{16} - \frac{80}{27} \frac{S_1^2}{M_0^4} - \frac{70}{9} \frac{M_2}{M_0^3} \right) v^8 \nn \\
& & + \left( \frac{81}{2} \frac{S_1}{M_0^2} + 6 \frac{S_1 M_2}{M_0^5} - 6 \frac{S_3}{M_0^4} \right) v^9 - \left( \frac{6615}{128} - \frac{115}{18} \frac{S_1^2}{M_0^4}\right.\nn\\
&&\left.- \frac{935}{24} \frac{M_2}{M_0^3} - \frac{35}{12} \frac{M_2^2}{M_0^6} + \frac{35}{12} \frac{M_4}{M_0^5} \right) v^{10} 
\nn \\
& & + \left( 165 \frac{S_1}{M_0^2}+ \frac{1408}{243} \frac{S_1^3}{M_0^6} + \frac{968}{27} \frac{S_1 M_2}{M_0^5} - \frac{352}{9}\frac{S_3}{M_0^4} \right) v^{11} \nn \\
& &  - \left( \frac{45927}{256} + \frac{123}{14} \frac{S_1^2}{M_0^4} - \frac{9147}{56} \frac{M_2}{M_0^3} - \frac{93}{4} \frac{M_2^2}{M_0^6}\right.\nn\\
&&\left.-24 \frac{S_1^2 M_2}{M_0^7} + 24 \frac{S_1 S_3}{M_0^6} + \frac{99}{4} \frac{M_4}{M_0^5} \right) v^{12} 
+ \mathcal{O} \left( v^{13} \right)
 \,, \label{DeltaERyan}
\end{eqnarray}
where $v\equiv (M_0 \Omega)^{1/3}$ is the linear velocity, $M_\ell$ are the Geroch-Hansen mass multipole moments, whereas $S_\ell$ are the Geroch-Hansen current multipole moments~\cite{Geroch:1970cd,Hansen:1974zz}. Note that these moments are equivalent~\cite{Gursel} to the multipole moments defined by Thorne~\cite{Thorne:1980ru} using asymptotically mass-centered Cartesian coordinates, and that the definitions above are all gauge invariant.

For the asymptotically-flat, stationary and axisymmetric spacetime discussed here and written as $ds^2=g_{\mu\nu}dx^\mu dx^\nu$, we have
\begin{eqnarray}
\Omega &=& \frac{-g_{t\varphi,r} + \sqrt{(g_{t\varphi,r})^2-g_{tt,r} g_{\varphi\varphi,r}}}{g_{\varphi\varphi,r}}\,, \label{Omega}\\
E_p &=& - \frac{g_{tt}+g_{t\varphi} \Omega}{\sqrt{-g_{tt}-2 g_{t\varphi} \Omega - g_{\varphi\varphi} \Omega^2}}\,. \label{Ep}
\end{eqnarray}

By using the definitions above and inverting the function $v=v(r)$, we obtain the expression of $\Delta E$ given in Appendix~\ref{app:DeltaE}. 
By comparing Eq.~\eqref{DeltaE} with Eq.~\eqref{DeltaERyan}, we can identify the nonvanishing multipole moments to quadratic order in the spin:
\begin{eqnarray}
 \frac{M_0}{M} &=& 1+{\delta m} \chi ^2-\frac{\gamma_0+12 \sqrt{5} \gamma  {\delta q} }{4 \sqrt{\pi }}\chi ^2\,, \label{moment0}\\
 \frac{S_1}{M^2} &=& \left(1 +\frac{\sqrt{3} \gamma_1-2 \sqrt{5} \gamma   }{4 \sqrt{\pi }}\right)\chi \,, \label{moment1}\\ 
 \frac{M_2}{M^3} &=& \left(\frac{4}{5} {\delta q}-1\right) \chi ^2-\frac{2 \gamma }{\sqrt{5 \pi }}-\frac{\chi ^2}{280 \sqrt{\pi }}\left[195 \sqrt{7} \gamma_3 
  \right.\nn\\ &&\left.
 +4 \sqrt{5} (28
   \gamma_2+\gamma  (135+56 {\delta m}+76 {\delta q}))\right]\,, \label{moment2}\\ 
 \frac{S_3}{M^4} &=& -\frac{3 \sqrt{7} {\gamma_3}+44 \sqrt{5} \gamma}{28 \sqrt{\pi }} \chi \,, \label{moment3}\\ 
 \frac{M_4}{M^5} &=& \frac{65 \sqrt{7} {\gamma_3}+4 \left(420 {\gamma_4}-\sqrt{5} \gamma  (221+432 {\delta q})\right)}{2940   \sqrt{\pi }} \chi^2\,, \nn\\\label{moment4}
\end{eqnarray}
Note that, because the expansion~\eqref{DeltaERyan} contains more coefficients than multipole moments, the comparison with Eq.~\eqref{DeltaE} is also a nontrivial consistency check of our solution, because some relations between the coefficients of the small$-v$ expansion are fixed through Einstein's vacuum equations.

Interestingly --~due to the coupling to $L=3$ axial and $L=4$ polar terms discussed in Sec.~\ref{sec:slowrot}~-- the
tidal field introduces a nonvanishing current octupole $S_3$ and a nonvanishing mass hexadecapole $M_4$ even if the
background solution does not possess such moments to second order in the spin. Clearly, such corrections would add to
the (spin-induced) $S_3$ and $M_4$ terms appearing in the background solution to third and to fourth order in the spin,
respectively~\cite{Yagi:2014bxa}. However, since we neglect terms higher than second
order in the spin, $S_3$ and $M_4$ are absent from our background solution, and only appear in the tidally-induced
deformation. This is similar to the nonspinning case, in which the tidal field induces a quadrupole moment [the first
  term in Eq.~\eqref{moment2}] even if the central object was originally spherically symmetric.

The result above provides a practical definition of the tidally-induced multipole moments of the spacetime. Once the exterior metric is matched to the interior solution, the constants $M$, $\chi$, $\delta m$, $\delta q$, $\gamma$ and $\gamma_\ell$ can be extracted and can be related to the multipole moments of the external spacetime using the definitions above. 

\subsection{Note on the definition of multipole moments}\label{sec:notemoments}
In Sec.~\ref{subsec:separation} we have discussed some subtleties and some degree of arbitrariness in separating the solution describing the external tidal field from that describing the response of the system, even at the linearized level. 
We wish here to comment on some further technical issue which is related to the definition of the multipole moments.

Our procedure is based on the fact that the perturbation equations are linear and, therefore, the two solutions mentioned above independently solve Einstein's vacuum field equations.
However, even if the full perturbed solution behaves --~by definition~-- linearly, the multipole moments of the spacetime might be mixed among the two solutions. In other words, the multipole moments of the central object might in principle be contaminated by the external solution. An example of this fact is the static and axisymmetric Weyl solution, whose line element reads
\begin{equation}
 ds^2=-e^{2U}dt^2+e^{2(k-U)}(d\rho^2+dz^2)+W^2 e^{-2U}d\varphi^2 \,,
\end{equation}
where $U$, $k$ and $W$ depend only on $\rho$ and $z$.
It is easy to prove that Einstein's equations impose the potential $U$ to be a harmonic function in flat space, $\nabla^2 U=0$. Therefore, in this case
a linear combination of two solutions (say $U_1$ and $U_2$) is still a solution of the Laplace equation.
However, the resulting moments will in general be a \emph{nonlinear} combination of the moments of the individual solutions~\cite{Geroch:1970cd}.
To linear order, $U_1$ and $U_2$ can be considered as small perturbations and therefore nonlinear terms can be neglected. However, this example shows that a mixing between two independent solutions can occur in the computation of the multipole moments. 

For this reason, it is not clear whether the moments of the linear-response solution are the \emph{true} moments of the
deformed compact object. We stress that this limitation is not a prerogative of our approach. The same criticism equally applies to the static case (see Ref.~\cite{Damour:2009vw} for a discussion). Although our procedure is reasonable and fully equivalent to previous approaches, we believe that a more rigorous analysis is needed to solve this important issue, even in the static case. This would likely require a fifth order post-Newtonian expansion of the field equations for a binary system, which is currently not available for generic mass ratios, in order to determine the quantities which actually appear in the post-Newtonian gravitational waveforms. Indeed, the definition of the tidal Love numbers in a relativistic theory would remain mostly academic without a proper connection to observable quantities. 

Leaving these problems for future work, in the following we simply follow the standard procedure to define the tidal Love numbers.

\subsection{Tidal Love numbers of a spinning object}

It is clear from Eqs.~\eqref{moment0}--\eqref{moment4} that the external tidal field modifies various multipole moments of the spacetime. Because such corrections are necessarily linear in the tidal field ${\cal E}_0$, we define various tidal Love numbers $\lambda_\ell^{(M)}$ and $\lambda_\ell^{(S)}$ as follows (cf. Eq.~\eqref{defLove}): 
\begin{eqnarray}
 \lambda^{(M)}_0 &=& -\frac{{\partial_{{\cal E}_0}\!\gamma_0}+12 \sqrt{5} \partial_{{\cal E}_0}\!{\gamma}  \,{\delta q} }{4 \sqrt{\pi }}M \chi ^2	\,, \label{Love0}\\
 \lambda^{(S)}_1 &=& \frac{\sqrt{3} \partial_{{\cal E}_0}\!{\gamma_1}-2 \sqrt{5} \partial_{{\cal E}_0}\!{\gamma}  }{4 \sqrt{\pi }}M^2 \chi \,, \label{Love1}\\
 \lambda^{(M)}_2 &=& -\frac{2 \partial_{{\cal E}_0}\!{\gamma}  M^3}{\sqrt{5 \pi }}-\frac{M^3\chi ^2}{280 \sqrt{\pi }}\left[195 \sqrt{7} \partial_{{\cal E}_0}\!{\gamma_3} \right.\nn\\
 &&\left.+4 \sqrt{5} (28
   \partial_{{\cal E}_0}\!{\gamma_2}+\partial_{{\cal E}_0}\!{\gamma}  (135+56 {\delta m}+76 {\delta q}))\right]\,,\nonumber\\
 &&\label{Love2}\\
 \lambda^{(S)}_3 &=& -\frac{3 \sqrt{7} \partial_{{\cal E}_0}\!{\gamma_3}+44 \sqrt{5} \partial_{{\cal E}_0}\!{\gamma}  }{28 \sqrt{\pi }} M^4  \chi\,, \label{Love3}\\
 \lambda^{(M)}_4 &=& \frac{M^5 \chi ^2}{2940   \sqrt{\pi }} \left[65 \sqrt{7} \partial_{{\cal E}_0}\!{\gamma_3}\right.\\
 &&\left.-4 \left(420 \partial_{{\cal E}_0}\!{\gamma_4}+\sqrt{5} \partial_{{\cal E}_0}\!{\gamma}\,  (221+432 {\delta q})\right)\right] \,, \label{Love4}
\end{eqnarray}
where $\partial_{{\cal E}_0}$ denotes derivative with respect to ${\cal E}_0\propto \alpha$ [cf. Eq.~\eqref{E0}]. Because the quantities above are linear in the tidal field, the derivatives with respect to ${\cal E}_0$ are just numbers (see Sec.~\ref{subsec:extracting} below).
The first term in Eq.~\eqref{Love2} is proportional to the standard electric quadrupolar Love number~\eqref{LoveNumber} and it is the only term that does not depend on the spin. The other terms in Eq.~\eqref{Love2}, as well as Eqs.~\eqref{Love0}--\eqref{Love1} and Eqs.~\eqref{Love3}--\eqref{Love4}, are novel spin-induced corrections.

The tidally-induced corrections to the multipole moments are linear- (or higher-) order quantities in the spin.
For an electric quadrupolar tidal field, the mass quadrupole gets tidal-induced corrections at quadratic order in the spin, i.e. the corrections enter at the same order of the spin-induced quadrupole moment of the Kerr metric. On the other hand, both the current octupole $S_3$ and the mass hexadecapole $M_4$ get tidally-induced corrections which enter at \emph{lower} order in the spin than the spin-induced corrections, which would enter at ${\cal O}(\chi^3)$ and ${\cal O}(\chi^4)$, respectively~\cite{Yagi:2014bxa}. 

More precisely, to quadratic order in the spin, the tidal correction to $M_2$ is suppressed by a factor $\alpha\ll1$ relative to the undeformed spin-induced term. However, the tidal corrections to $S_3$ and $M_4$ are dominant with respect to higher-order spin-induced corrections~\cite{Yagi:2014bxa} whenever $\chi^2\lesssim\alpha$. This condition is consistent with our perturbative scheme because both $\alpha$ and $\chi$ are small perturbation parameters, although it might be difficult to match in practice, except for extremely slow rotations which would however make the spin-induced tidal Love numbers~\eqref{Love0}--\eqref{Love4} almost zero.

Finally, here we focused on the most interesting case of the tidal Love numbers associated with a quadrupolar electric tidal field. Nonetheless, there is no reason to expect qualitatively different results for other components of the tidal field. The selection rules discussed in Sec.~\ref{sec:slowrot} suggest that an axisymmetric electric tidal field with  multipole $\ell$ would introduce corrections to the mass multipole moment $M_\ell$ to ${\cal O}(\chi^0)$ and ${\cal O}(\chi^2)$, to the mass multipole moments $M_{\ell\pm2}$ to ${\cal O}(\chi^2)$, and would introduce corrections to the current multipole moments $S_{\ell\pm1}$ to ${\cal O}(\chi)$. Likewise, we expect that a magnetic tidal field with multipole $\ell$ would introduce corrections to the current multipole moment $S_\ell$ to ${\cal O}(\chi^0)$ and ${\cal O}(\chi^2)$, would modify the moments $S_{\ell\pm2}$ to ${\cal O}(\chi^2)$, and would introduce corrections to the mass multipole moments $M_{\ell\pm1}$ to ${\cal O}(\chi)$. A natural extension of our work is to compute these novel families of Love numbers\footnote{Note that an electric (resp. magnetic) tidal field with odd (resp. even) values of $\ell$ would generate perturbations which break the reflection symmetry of the background~\eqref{metric}. Consistently with the argument just presented, such components of the tidal field would induce multipoles such as $M_3$, $S_2$, etc.., which are identically vanishing in the case of a solution with reflection symmetry as the one discussed here.}.

In particular, an axisymmetric magnetic tidal field with $\ell=3$ should modify the mass quadrupole moment $M_2$ by a term linear in the spin and linear in the intensity of the tidal field. 
Because such term enters at \emph{linear} order in the spin, it might be the dominant deformation of the mass quadrupole $M_2$, although it would be suppressed by the fact that the octupolar magnetic component of the tidal field is much smaller than the quadrupolar electric component for typical sources.

\subsection{Extracting the Love numbers of a spinning object} \label{subsec:extracting}
In general, computing the Love numbers~\eqref{Love0}--\eqref{Love4} requires a numerical integration of the perturbation equations in the interior of the central object and a matching procedure with the analytical exterior solution discussed here. 

If the central object is a self-gravitating fluid, the interior solution consists of the metric perturbations and the fluid perturbations, the latter vanishing in the exterior.
By requiring regularity at the center of the object and continuity at the surface, the Love number~\eqref{LoveNumber} in the static case can be extracted~\cite{Hinderer:2007mb} from the ratio ${H_0^{(2)}}'/H_0^{(2)}$ evaluated at the radius $R$ of the star and by using the analytical solution~\eqref{polL2spin0}. Likewise, the constants $\gamma_\ell$ ($\ell=0,1,..,4$) can be extracted from $\delta H_0^{(0)}(R)$, $h_0^{(1)}(R)$, $\delta H_0^{(2)}(R)$, $h_0^{(3)}(R)$ and $\delta H_0^{(4)}(R)$, respectively, and using the analytical solution presented in this work.
After this matching, the parameters $\gamma$ and $\gamma_\ell$ will be necessarily proportional to $\alpha\propto{\cal E}_0$ and, therefore, the dependence on ${\cal E}_0$ of the Love numbers~\eqref{Love0}--\eqref{Love4} disappears, as expected. One can conveniently factorize the ${\cal E}_0$-dependence in the constants, by defining say $\gamma_0\equiv {\delta m}_{\rm tidal} {\cal E}_0$, and the matching procedure would then allow us to extract $ {\delta m}_{\rm tidal}$, which represents a mass shift induced by the tidal field at quadratic order in the spin, similarly to the parameter $\delta m$ that represents a mass shift purely induced by the angular momentum of the background solution. Similarly, the tidal Love numbers~\eqref{Love0}--\eqref{Love4} also depend on the coupling between the static Love number $k_{\rm el}^{(2)}\propto \gamma/\alpha$ and the quadrupole shift $\delta q$, which also appears at ${\cal O}(\alpha \chi^2)$.

In a subsequent work, we will apply this procedure to compute the tidal Love numbers~\eqref{Love0}--\eqref{Love4} for a spinning neutron star. 
In the next section, we will instead focus on the case in which the central object is a BH, which can be remarkably solved analytically.

\section{Tidally-distorted spinning BHs} \label{sec:BH}
The solution discussed above becomes much more tractable in the BH case. The latter can be obtained by requiring
regularity of the solution across and outside the event horizon. Because the Hartle-Thorne coordinates~\cite{Hartle:1967he,Hartle:1968si} in which our ansatz~\eqref{metric} is written are singular at $r=2M$, the regularity of the metric perturbations is slightly more subtle than in a regular set of coordinates as, e.g., that adopted in Ref.~\cite{Poisson:2014gka}. Nonetheless, to ensure regularity one can compute some curvature invariant, such as the Kretschmann scalar $R_{abcd}R^{abcd}$ and the Pontryagin density ${}^{*}\!RR\equiv\frac{1}{2}\epsilon^{abef}R_{abcd}R^{cd}{}_{ef}$, and impose regularity at the singular point.

For the background solution given in Appendix~\ref{app:background}, regularity imposes $\delta
q=0$ whereas $\delta m$ can be set to zero without loss of generality through a redefinition of the mass of the
background solution. Furthermore, as previously mentioned, regularity also imposes $\gamma=0$ in the
solution~\eqref{polL2spin0}--\eqref{K0inf}. It is also easy to show that the curvature invariants are divergent at $r=2M$ unless
\begin{equation}
  \gamma_2=\gamma_3=\gamma_4=0\,. \label{constr}
\end{equation}
Due to the parity properties of our solution, regularity of the Kretschmann scalar fixes the constants $\gamma_2$ and $\gamma_4$ related to the even-parity (polar) perturbations, whereas regularity of Pontryagin density fixes the constant $\gamma_3$ associated with the odd-parity (axial) perturbations.
As shown in Eqs.~\eqref{moment0} and \eqref{moment1}, the constants $\gamma_0$ and $\gamma_1$ are related to tidally-induced mass and spin shifts, respectively, and can be set to zero without loss of generality in the BH case, because they can be reabsorbed in the definition of mass and spin of the background Kerr metric. After imposing $\gamma_0=\gamma_1=0$ and the conditions~\eqref{constr}, the Kretschmann scalar reads
 \begin{eqnarray}
 &&M^4 R_{abcd}R^{abcd}= \frac{48}{y^6}+\sqrt{\frac{5}{\pi }}\frac{18  \alpha}{y^3} \left[1+3 \cos(2\vartheta)\right]\nn\\
 &&+\frac{72 \chi ^2}{y^{10}} \left(2+y-6 y^2-\left(8 y^2+y-6\right)
   \cos(2\vartheta)\right)\nn\\
   &&+\frac{27 \alpha  \chi ^2}{4 \sqrt{5 \pi } y^{10}}\left[(104+y (112-y (856 \right.\nn\\
   &&\left.+y (783+y (45 y-311))))-4 (104+y (-104\right.\nn\\
   &&\left.+y (48+5 y (79+5 (y-1)
   y)))) \cos(2\vartheta)\right.\nn\\
   &&\left.-5 (40+y (y (y (65+y (31+35 y))-56)-48)) \cos(4\vartheta)\right]\,,\nn
\end{eqnarray}
 and is regular everywhere except at $r=0$. A similar regular expression can be obtained for the Pontryagin density. Note that the Kretschmann scalar can also be decomposed in spherical harmonics as $M^4 R_{abcd}R^{abcd}= \sum_{i=0}^2 f_{2i}(r) Y^{2i0}(\vartheta)$, where $f_{0,2,4}(r)$ are radial functions. The exact regular metric of a tidally distorted spinning BH to second order in the spin is given in Appendix~\ref{app:BHsol}.

As a consistency check, we have verified that, to first order in the spin, our solution reduced
to that found by Poisson in Ref.~\cite{Poisson:2014gka} in the axisymmetric case with zero magnetic quadrupolar component of the tidal field. The transformation between our coordinate system and the null
coordinates of Ref.~\cite{Poisson:2014gka} is given in Appendix~\ref{app:solutions},
Eqs.~(\ref{trasft})-(\ref{trasfph}).

\subsection{Love numbers of a Kerr BH} \label{sec:LoveBH}
We can now compute the tidal Love numbers for a spinning BH. By imposing regularity through $\delta q=\gamma=0$, using Eq.~\eqref{constr}, and setting $\delta m=\gamma_0=\gamma_1=0$, from Eqs.~\eqref{moment0}--\eqref{moment4} we obtain

\begin{eqnarray}
 \frac{\partial M_2}{\partial {\cal E}_0} &=& \frac{\partial S_3}{\partial {\cal E}_0}=\frac{\partial M_4}{\partial {\cal E}_0}=0\,, \
\end{eqnarray}
together with the equations $\frac{\partial M_0}{\partial\alpha}=\frac{\partial S_1}{\partial\alpha}=0$, which just represent the freedom of rescaling the mass and spin of the background metric. Therefore, we immediately obtain that the Love numbers of a Kerr BH are zero as in the Schwarzschild case~\cite{Binnington:2009bb,Guerlebeck:2015xpa} and as recently found to first order in the spin by Landry and Poisson~\cite{Poisson}. We note that the separation of solutions discussed in Sec.~\ref{subsec:separation} is crucial to obtain such result. Any other prescription would account for a shift as in Eq.~\eqref{shift} which would modify the Love numbers. Therefore, the prescription given above is the unique one that yields zeroth Love numbers for a Kerr BH to second order in the spin. On the light of these results, it is also natural to conjecture that the Love numbers of a Kerr BH are zero to \emph{any} order, at least in the axisymmetric case considered here.

As a by-product of the BH uniqueness and no-hair
theorems~\cite{Carter71,Hawking:1973uf} (see also \cite{Heusler:1998ua,Chrusciel:2012jk,Robinson}),
the multipole moments of any stationary BH in isolation can be written
as~\cite{Hansen:1974zz},
\begin{equation}
 M_\ell+i S_\ell =M^{\ell+1}\left(i\chi\right)^\ell\,. \label{nohair}
\end{equation}
All moments with $\ell\geq2$ can be written in terms of $M_0=M$ and $S_1=J$ through the
above relation. Therefore, any independent measurement of three multipole moments (e.g. the mass, the spin and the mass quadrupole
$M_2$) is a null-test of the Kerr metric and, in turn, it might provide the first genuine strong-gravity confirmation of
general relativity~\cite{Psaltis:2008bb,Gair:2012nm,Yunes:2013dva,Berti:2015itd}.
Our results (together with those of Ref.~\cite{Poisson}) show that the no-hair relations~\eqref{nohair} are robust not only to nonperturbative effects in the tidal field~\cite{Guerlebeck:2015xpa}, but also to perturbative tidal-spin interactions. In other words, the relation~\eqref{nohair} holds also for a slowly-rotating
BH immersed in a weak tidal field and it can therefore be interpreted as a generalization of the no-hair theorems for
stationary, tidally-deformed spinning BHs.

\subsection{Properties of the solution}
\begin{figure}[ht]
\begin{center}
\epsfig{file=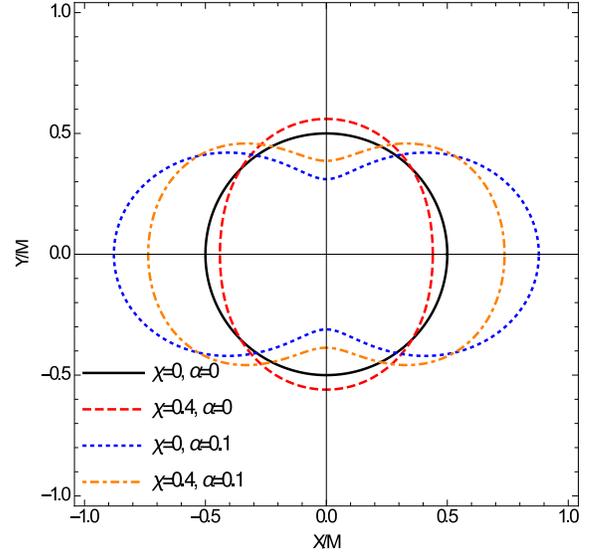,width=7.5cm,angle=0,clip=true}
\caption{
The curvature of the intrinsic geometry of a tidally-deformed spinning BH for different values of the spin $\chi$ and of the intensity of the tidal field $\alpha$. We used extreme values of $\chi$ and $\alpha$ in order to magnify the effect of the deformations. The coordinates $(X,Y)$ are related to $(r,\vartheta)$ through $r=\sqrt{X^2+Y^2}$, $Y/X=\tan^{-1}\vartheta$. 
\label{fig:intrinsicR}}
\end{center}
\end{figure}

The BH solution given in Appendix~\ref{app:BHsol} is an analytical (albeit perturbative) solution of Einstein's vacuum equations and, as such, it is interesting {\it per se}. It is therefore relevant to study this solution more in detail, e.g. by computing various geometrical and geodesic quantities related to this spacetime. Let us start by computing the intrinsic geometry at the horizon. The horizon location is defined as the largest root of $g^{rr}$ and reads
\begin{equation}
 r_+=2M\left[1-\frac{\chi^2}{8}-\frac{3\alpha\chi^2}{2\sqrt{5\pi}}\right]\,,\label{horizon}
\end{equation}
Although the horizon location does not depend on the coordinates $\vartheta$ and $\varphi$, its intrinsic geometry is not spherical. To compute the intrinsic metric, we consider the spatial section $dt=0$ of the metric~\eqref{metric} at $r=r_+$. In the slowly-rotating limit, we obtain
\begin{equation}
 ds^2_{t={\rm const}, r=r_+}=g_{\vartheta\vartheta}(r=r_+,\vartheta)d\Omega^2\,.
\end{equation}
and therefore the intrinsic geometry is spherical only when $g_{\vartheta\vartheta}$ evaluated at $r=r_+$ does not depend on $\vartheta$. Using the solution~\eqref{BHgthethatheta}, the Ricci curvature of the intrinsic geometry reads
\begin{eqnarray}
 &&M^2 R_{\rm intr} = \frac{1}{2}-\frac{3}{8} \chi ^2 \cos(2\vartheta)+\frac{3}{4} \sqrt{\frac{5}{\pi }} \alpha \left[1+3 \cos(2\vartheta)\right]\nn\\
 &&-\frac{9 \alpha    \chi ^2}{128 \sqrt{5 \pi }} \left[1+172 \cos(2\vartheta)+115 \cos(4\vartheta)\right]\,,
\end{eqnarray}
and is constant only in the nonrotating and undeformed case. The intrinsic curvature is shown in Fig.~\ref{fig:intrinsicR} for different values of $\chi$ and $\alpha$.

It is also interesting to look at the geodesic structure of the spacetime. 
For a stationary and axisymmetric spacetime, one can define the potential
\begin{equation} 
V(r)\equiv-g_{rr}^{-1}E_p^2\left[E_p^2{\cal U}(r,\pi/2)+1\right]\,,
\end{equation}
where ${\cal U}(r,\vartheta)=g^{tt}-2lg^{t\varphi}+l^2g^{\varphi\varphi}$,
$E_p$ is the energy per unit mass of a point particle (given in Eq.~\eqref{Ep}), and $l$ is the proper angular momentum. 
For a circular, equatorial orbit at $r=r_c$, $E_p$ and $l$ can be determined by
imposing $V(r_c)=V'(r_c)=0$, where the prime indicates
differentiation with respect to $r$. The solution for $E_p$ is given in Eq.~\eqref{Ep}.  The further condition $V''=0$ yields the location of the innermost stable circular orbit (ISCO), $r=r_{\rm ISCO}$. 

Furthermore, by considering small perturbations of a circular, equatorial orbit, one finds the epicyclic frequencies $\Omega_r$ and $\Omega_\vartheta$ governing small oscillations in the radial and in the $\vartheta$ direction, respectively. These read
\begin{eqnarray} 
\Omega_r^2&=&\frac{(g_{tt}+\Omega
g_{t\varphi})^2}{2g_{rr}}\frac{\partial^2{\cal U}}{\partial
r^2}\left(r_c,\frac{\pi}{2}\right)\,,\label{Omegar}\\
\Omega_\vartheta^2&=&\frac{(g_{tt}+\Omega
g_{t\varphi})^2}{2g_{\vartheta\vartheta}}\frac{\partial^2{\cal
U}}{\partial\vartheta^2} \left(r_c,\frac{\pi}{2}\right)\,.\label{Omegatheta}
\end{eqnarray}
Using Eqs.~\eqref{Omega},~\eqref{Omegar} and~\eqref{Omegatheta}, it is straightforward to show that, in our case, the explicit form of the orbital and epicyclic frequencies reads
\begin{eqnarray}
 &&M \Omega = \frac{1}{y^{3/2}}+\sqrt{\frac{5}{\pi }}\frac{3 \left(y^3-2\right) \alpha }{8 y^{3/2}}-\frac{\chi }{y^3}+\frac{9 \alpha  \chi }{2 \sqrt{5\pi } y^3}\nn\\
 &&+\frac{(3+y) (3 y-2) \chi ^2}{4 y^{11/2}}\nn\\
 &&+\frac{3 (180+y (y (y (177+(101-15 y) y)-174)-110)) \alpha  \chi ^2}{32 \sqrt{5 \pi
   } y^{11/2}}\,,\nn\\
   \end{eqnarray}
   \begin{eqnarray}
 &&(M \Omega_r)^2 = \frac{y-6}{y^4}+\sqrt{\frac{5}{\pi}}\frac{3  (y-2) \left(-2-13 y+4 y^2\right) \alpha }{4 y^3}\nn\\
 &&+\frac{6 (2+y) \chi }{y^{11/2}}+\frac{3 \left(-12+34
   y-240 y^2+120 y^3+15 y^4\right) \alpha  \chi }{4 \sqrt{5 \pi } y^{11/2}}\nn\\
   &&+\frac{\left(48-66 y-47 y^2-3 y^3\right) \chi ^2}{2 y^8}\nn\\
   &&-\frac{3
   \left(960-216 y+8 y^2-924 y^3+675 y^4+434 y^5\right) \alpha  \chi ^2}{16 \sqrt{5 \pi } y^8} \,,\nn\\
      \end{eqnarray}
   \begin{eqnarray}
 &&(M \Omega_\vartheta)^2 = \frac{1}{y^3}-\sqrt{\frac{5}{\pi}}\frac{3 (y-2)^2 (2y-1) \alpha }{4 y^3}-\frac{6 \chi }{y^{9/2}}\nn\\
 &&-\frac{3 \left(74-120 y+30 y^2+15 y^3\right)
   \alpha  \chi }{4 \sqrt{5 \pi } y^{9/2}}+\frac{\left(-6+25 y+9 y^2\right) \chi ^2}{2 y^7}\nn\\
   &&+\frac{3 \left(-120+1048 y-948 y^2+57 y^3+346 y^4\right)
   \alpha  \chi ^2}{16 \sqrt{5 \pi } y^7}\,.
\end{eqnarray}
By evaluating these frequencies at the ISCO, we obtain
\begin{eqnarray}
 M \Omega^{\rm ISCO}&=& \frac{1}{6 \sqrt{6}}+\frac{491}{8} \sqrt{\frac{5}{6 \pi }} \alpha +\left(\frac{11}{216}-\frac{6671 \alpha }{48 \sqrt{5 \pi }}\right) \chi \nn\\
 &&+\left(\frac{59}{648 \sqrt{6}}+\frac{5819}{864}    \sqrt{\frac{5}{6 \pi }} \alpha \right) \chi ^2 \,, \\ \label{OmegaISCO}
  M \Omega_\vartheta^{\rm ISCO}&=& \frac{1}{6 \sqrt{6}}+ 37 \sqrt{\frac{5}{6 \pi }}\alpha +\left(\frac{1}{24}-\frac{35209 \alpha }{288 \sqrt{5 \pi }}\right) \chi \nn\\
  &&+\left(\frac{57539 \alpha }{864 \sqrt{30 \pi }}+\frac{79}{1296 \sqrt{6}}\right) \chi ^2 \,,  \label{OmegathetaISCO}
\end{eqnarray}
whereas the ISCO radial epicyclic frequency $\Omega_r^{\rm ISCO}$ vanishes to second order in the spin, as in the undeformed Kerr case.

\subsection{Tidal versus spin effects}
Our results can be used to estimate the effects of spin and tidal deformations on the BH geometry. The analysis of this section will be mostly qualitative, a more detailed study will appear elsewhere. Let us assume that the tidal field is generated by a source of mass $m_c$ at a distance $r_0$ from the central BH\footnote{We recall that we are considering an axisymmetric tidal field. Therefore, strictly speaking, our source should be a ring of mass $m_c$ and radius $r_0$. Nonetheless, the qualitative results of this section would also apply to the more realistic case in which the source is a companion star of mass $m_c$ at orbital distance $r_0$. In this case ${\cal E}_{m\neq0}\neq0$ and the tidal field sources \emph{both} axisymmetric and nonaxisymmetric deformations in the metric. By virtue of the axisymmetry of the background, modes with different azimuthal number $m$ are decoupled from each other and the axisymmetric components of the metric perturbations are exactly described by our solution.
}. 
In this case, ${\cal E}_0\sim m_c/r_0^3$ and Eq.~\eqref{E0} yields
\begin{equation}
 \alpha\sim \frac{m_c M^2}{r_0^3}\,,\label{alpha}
\end{equation}
which is understood as an order of magnitude estimate. We also assume $m_c\sim M$ so that $\alpha\sim (M/r_0)^3$. In this case,
\begin{eqnarray}
 \frac{\Omega^{\rm ISCO}}{\Omega^{\rm ISCO}_0} &\approx& 1+0.150 \chi_{0.2}+0.022 \chi_{0.2}^2+0.017 r_{30}^{-3}\nn\\
 &&-0.004 r_{30}^{-3} \chi_{0.2}+0.00008 r_{30}^{-3} \chi_{0.2}^2 \,,\\
 \frac{\Omega^{\rm ISCO}_\vartheta}{\Omega^{\rm ISCO}_{\vartheta,0}} &\approx&  1+0.123 \chi_{0.2}+0.015 \chi_{0.2}^2+0.010 r_{30}^{-3}\nn\\
 &&-0.003 r_{30}^{-3} \chi_{0.2}-0.0002 r_{30}^{-3} \chi_{0.2}^2\,, 
\end{eqnarray}
where each denominator denotes the corresponding frequency evaluated at $\chi=\alpha=0$, $\chi_{0.2}=\chi/(0.2)$ and $r_{30}=r_0/(30M)$. As expected, tidal corrections are small because the tidal field is suppressed by the third power of the orbital distance. However, for $r_0\approx 30 M$ (i.e., $\alpha\sim 10^{-5}$) and $\chi\approx 0.2$, the linear tidal correction is comparable to the quadratic in spin correction and they both correct the static, undeformed result by a few percent. 

With the normalization of $\chi$ and $r_0$ adopted above, the correction linear in the spin and in the tidal field --~which arises from the spin-tidal coupling~-- is of the order of $0.5\%$, whereas the correction quadratic in the spin and linear in the tidal field is much smaller, of the order of $0.02\%$. 
Note, however, that our metric is valid in the limit $r\ll r_0$, whereas the frequencies above are evaluated at the ISCO, $r_{\rm ISCO}\sim 6M$, which is only moderately smaller than $r_0\sim 30M$. For this reason, our discussion is intended only as an order-of-magnitude estimate.

\section{Conclusions and extensions} \label{sec:conclusions}
Computing the tidal Love numbers of a spinning neutron star is an open problem with various potential applications in classical general relativity and in gravitational-wave astronomy.
Here, we presented a framework to study gravitational perturbations of a slowly-rotating geometry to second order in the spin. We applied this technique to study static, axisymmetric tidal perturbations of a spinning object. Remarkably, the perturbation equations can be solved analytically in vacuum. We provided the explicit form of the metric describing the exterior geometry of a spinning object distorted by an axisymmetric tidal field to second order in the spin. 

Because of spin couplings, an external quadrupolar electric tidal field deforms the quadrupole moment $M_2$ of the central object up to quadratic order in the spin, the dipole and octupole current moments $S_1$ and $S_3$ to linear order in the spin, and also deforms the monopole and hexadecapole mass moments $M_0$ and $M_4$ to quadratic order in the spin. Correspondingly, for a spinning object a new class of different Love numbers emerges, while the standard Love numbers acquire spin-induced corrections. 

When the central object is a spinning BH, the metric simplifies considerably. 
Similarly to the nonrotating case, we have shown that the multipole moments of a Kerr BH are not affected by the tidal field at least up to quadratic order in the spin, and thus the corresponding Love numbers are zero. 
This implies that the no-hair relation~\eqref{nohair} is not affected by perturbative spin-tidal interactions.
This result is strongly based on the discrimination between the solution describing the tidal field and that describing the linear response of the central object, which we have discussed in some detail. Furthermore, we provided the metric describing a tidally-deformed Kerr BH in concise form to second order in the spin and computed various physical properties of the solution, including the epicyclic frequencies and the intrinsic geometry of the event horizon.

These findings have potentially important implications for gravitational-wave phenomenology with ground-based detectors~\cite{aLIGO,aVIRGO,KAGRA}. For example, current gravitational-wave templates for compact binary inspirals adopt Love numbers which are valid for nonspinning objects, neglecting the spin corrections introduced here. In this context, it is crucial to understand how the rotational Love numbers defined here enter the gravitational waveforms, similarly to what has been done in the past for the nonrotating case~\cite{Flanagan:2007ix,Hinderer:2009ca,Baiotti:2010xh,Baiotti:2011am,Vines:2011ud,Pannarale:2011pk,Lackey:2011vz,Damour:2012yf,Vines:2010ca,Lackey:2013axa,Favata:2013rwa,Yagi:2013baa}.

Our deformed BH solution might be of interest for null tests of the Kerr geometry and tests of general relativity based on various observations, e.g. tracing of BH shadows with the Event Horizon Telescope~\cite{Loeb:2013lfa}, detection of quasiperiodic oscillations in the signal emitted by accreting BHs with the X-ray telescope LOFT~\cite{Feroci:2011jc,Maselli:2014fca}, and observations of gravitational-wave from extreme-mass ratio inspirals with eLISA~\cite{AmaroSeoane:2012je,Yunes:2013dva,Berti:2015itd}. A more detailed characterization of the solution and its phenomenological applications are left for future work.

In a subsequent work~\cite{future}, we plan to use the results presented here to compute the Love numbers of a spinning neutron star explicitly. This requires solving the perturbation equations in the interior of the neutron star, taking into account also fluid perturbations and matching the interior solution with the exterior metric presented in this paper.
Another natural question we wish to answer is whether the spin-induced corrections to the tidal Love numbers of a neutron star satisfy some nearly-universal relations as their static counterpart~\cite{Yagi:2013bca,Yagi:2013awa}.

In this work, we focused on the most relevant, polar-led $L=2$ perturbations, i.e. we assumed that the tidal field has only a quadrupolar electric component at the leading order. Extending our results to the axial-led sector~\eqref{axial_led} (thus including a magnetic component of the tidal field) and to other values of $\ell$ is straightforward. As explained in the end of Sec.~\ref{sec:LoveBH}, novel families of Love numbers will emerge and they would correct the multipolar structure of a spinning neutron star or of a spinning BH in agreement with the selection rules discussed in Sec.~\ref{sec:slowrot}.

Indeed, the corrections to the multipole moments of a tidally-distorted spinning neutron star will modify the approximate three-hair relations that exist for isolated compact stars~\cite{Pappas:2013naa,Stein:2014wpa,Yagi:2014bxa}. For example, the mass quadrupole moment $M_2$ would acquire ${\cal O}(\chi)$ corrections proportional to the $\ell=3$ magnetic component of the external tidal field and would acquire ${\cal O}(\chi^2)$ corrections proportional to the $\ell=4$ electric component of the external tidal field. Higher multipole moments would also be modified accordingly to the selection rules discussed in Sec.~\ref{sec:slowrot}. Such corrections will be explicitly presented in a forthcoming publication~\cite{future}.

Although our analysis extends part of the results of Ref.~\cite{Poisson:2014gka} to quadratic order in the spin, we did not attempt to identify the tidal moments of the source. It would be interesting to complement our analysis by matching the exterior solution presented here (which is valid in the region $R<r\ll r_0$) to a post-Newtonian metric describing the source of the tidal field in an overlapping region~\cite{Poisson:2014gka}. This would allow us to express the constant $\alpha$ in terms of the source parameters up to a certain post-Newtonian order.

A major limitation of our results is the assumption of axisymmetry, i.e. we restricted to $m=0$ tidal perturbations. The tidal field produced by an orbiting companion is not axisymmetric, so our results describe only the axisymmetric part\footnote{It is important to stress that the Love numbers do not depend on the details of the source. Therefore, our results describe the (axisymmetric, quadrupolar) rotational Love numbers of a spinning body immersed in a generic tidal field, regardless of the symmetry of the latter.} of such tidal interaction (the axisymmetry of the background guarantees that modes with different $m$ are decoupled from each other). 
To second order in the spin, a nonaxisymmetric tidal field would induce precession of the  central object's angular momentum, introducing time dependence in the problem. This effect can be avoided by restricting the analysis to first order in the spin, as done in Refs.~\cite{Poisson:2014gka,Poisson}. 
In this case our method can be straightforwardly extended to compute the new Love numbers in the $m\neq0$ case to linear order in the spin.
In principle, however, our method is also well suited to deal with time-dependent perturbations and we hope to come back to this interesting generalization in the future.
Indeed, the slow-rotation framework presented in this paper can be applied to the case of time-dependent perturbations with only minor modifications~\cite{Pani:2013pma}. Such extension can be also relevant for a variety of important problems, including the computations of the quasinormal modes of spinning neutron stars~\cite{Kokkotas:1999bd,Ferrari:2007dd} to second order in the spin.

\noindent{\bf{\em Note added.}}
After this work was in its final stages, we were informed that Poisson and Landry are extending~\cite{Poisson} the results of Ref.~\cite{Poisson:2014gka} to the case of a material object and of nonaxisymmetric tidal perturbations. Our formalism and theirs are in various ways complementary to each other: differently from Ref.~\cite{Poisson} we extended the perturbative analysis to second order in the spin, but we restricted to the axisymmetric case and did not include a quadrupolar magnetic component of the tidal field.

\begin{acknowledgments}
We are particularly indebted to Eric Poisson and Phil Landry for useful discussions and comments. We also thank
  Norman G{\"u}erlebeck, Tanja Hinderer and Jan Steinhoff for interesting discussions.
We also thank an anonymous referee for useful comments on the definition of the multipole moments in tidally-deformed spacetimes.
P.P. was supported by the European Community through
the Intra-European Marie Curie Contract No.~AstroGRAphy-2013-623439 and by FCT-Portugal through the project IF/00293/2013.
This work was partially supported by ``NewCompstar'' (COST action MP1304) and by the NRHEP 295189 FP7-PEOPLE-2011-IRSES
Grant. Some of the computations were performed at the cluster {\it Baltasar-Sete-S\'{o}is} supported by the DyBHo-256667
ERC Starting Grant.
\end{acknowledgments}

\appendix
\begin{widetext}

\section{Background geometry to second order in the spin}\label{app:background}
Here we give the metric coefficients entering the background geometry~\eqref{metric} to second order in the spin~\cite{Hartle:1967he,Hartle:1968si}:
\begin{eqnarray}
 && {\cal M}(r)=M\,,	\quad e^{\nu(r)}=1-\frac{2M}{r}\,, \quad \omega(r)=\frac{2\chi M^2}{r^3}\,,\\
 && m_0(r)= \chi^2 M\left(\delta m  -\frac{ M^3}{r^3}\right)\,,\\
 && m_2(r)=-\chi^2\left(\frac{\delta m M}{r-2M}-\frac{M^4}{r^3 (r-2 M)}\right)\,,\\
 && m_2(r)= \frac{M^3 \chi ^2}{r^4} (5 M-r) (r-2 M)\nn\\
 &&+\frac{\delta q\chi ^2}{2 M^2 r}  \left(M (M-r) \left(3 r^2-2 M^2-6 M r\right)+3 r^2 (r-2 M)^2 \tanh ^{-1}\left(\frac{M}{r-M}\right)\right)\,,\\
 && j_2(r)= \frac{M^3 \chi ^2 (M+r)}{r^4}\nn\\
 &&+\frac{\delta q \chi ^2}{2 M^2 r (r-2 M)} \left(M (r-M) \left(3 r^2-2 M^2-6 M r\right)-3 r^2 (r-2 M)^2 \tanh ^{-1}\left(\frac{M}{r-M}\right)\right)\,,\\
 && v_2(r)=-\frac{M^4 \chi ^2}{r^4}+\frac{\delta q \chi ^2}{M} \left(3 (r-M) \tanh ^{-1}\left(\frac{M}{r-M}\right)-\frac{M \left(M^2+3 r^2-6 M r\right)}{r (r-2 M)}\right)\,,
\end{eqnarray}
the meaning of the various constants is explained in the main text. Note that we have factored out the spin dependence of the second-order terms, so that $\delta m$ and $\delta q$ are ${\cal O}(\chi^0)$ numbers which multiply terms proportional to $\chi^2$.
\section{Stationary perturbations of a slowly-rotating relativistic object}\label{app:Kojima}
In this Appendix we briefly present the derivation of the field equations; to reduce the risk of typographical errors and facilitate comparison
with our results, we made the entire calculation available online in the Supplemental Material.

As a background, we consider the spinning geometry~\eqref{metric} to second order in the rotation rate (cf. Appendix~\ref{app:background}) and
we consider a harmonic decomposition of the metric perturbations as
\begin{equation}
 \delta g_{\mu\nu}(t,r,\vartheta,\varphi)=\delta g_{\mu\nu}^{\rm odd}(t,r,\vartheta,\varphi)+\delta g_{\mu\nu}^{\rm even}(t	,r,\vartheta,\varphi)
\end{equation}
with
\begin{equation}\label{oddpart}
\delta g_{\mu\nu}^{\rm odd} =
 \begin{pmatrix}
  0 & 0 & h_0^\ell S_\vartheta^{\ell} & h_0^\ell S_\varphi^{\ell} \\
  * & 0 & h_1^\ell S_\vartheta^{\ell} & h_1^\ell S_\varphi^{\ell} \\
  *  & *  & -h_2^\ell\frac{X^\ell}{\sin\vartheta} & h_2^\ell\sin\vartheta W^\ell  \\
  * & * & * & h_2^\ell\sin\vartheta X^\ell
 \end{pmatrix}\,,
\end{equation}
\begin{equation}\label{evenpart}
\delta g_{\mu\nu}^{\rm even}=
\begin{pmatrix}
g_{tt}^{(0)} H_0^\ell Y^\ell & H_1^\ell Y^\ell & \eta_0^\ell Y_{,\vartheta}^\ell& \eta_0^\ell Y_{,\varphi}^\ell\\
  * & g_{rr}^{(0)} H_2^\ell Y^\ell & \eta_1^\ell Y_{,\vartheta}^\ell & \eta_1^\ell Y_{,\varphi}^\ell\\
  *  & *  & r^2\left[K^\ell Y^\ell+G^\ell W^\ell\right] & r^2  G^\ell X^\ell  \\
  * & * & * & r^2\sin^2\vartheta\left[K^\ell Y^\ell-G^\ell W^\ell\right]
 \end{pmatrix}\,,
\end{equation}
%
where asterisks represent symmetric components, $g_{tt}^{(0)}=e^\nu$, $1/g_{rr}^{(0)}-1-2{\cal M}/r$, $Y^{\ell}=Y^{\ell}(\vartheta,\varphi)$ are the scalar spherical harmonics and we have defined
\begin{eqnarray}
(S_\vartheta^{\ell},S_\varphi^{\ell})&\equiv&\left(-\frac{Y^{\ell}_{,\varphi}}{\sin\vartheta}
,\sin\vartheta Y^{\ell}_{,\vartheta}\right)\,.\\
(X^{\ell},W^{\ell})&\equiv&\left(2(Y^{\ell}_{,\vartheta\varphi}-\cot\vartheta Y^{\ell}_{,\varphi}),Y^{\ell}_{,\vartheta\vartheta}-\cot\vartheta Y^{\ell}_{,\vartheta}-\frac{Y^{\ell}_{,\varphi\varphi}}{\sin^2\vartheta}\right)\,, \label{XW}
\end{eqnarray}
which are related to the vector and tensor spherical harmonics, respectively.
Here and in the following, a sum over the harmonic indices $\ell$ and $m$ (such that $|m|\leq\ell$) is implicit\footnote{Furthermore, from now on we will append the relevant   multipolar index $\ell$ to any perturbation variable but we will omit the index $m$, because in an axisymmetric background it is possible to decouple the perturbation equations so that all quantities have the same value of $m$.}.
Under parity transformations ($\vartheta\rightarrow\pi-\vartheta$,
$\varphi\rightarrow\varphi+\pi$): polar and axial perturbations are
multiplied by $(-1)^\ell$ and $(-1)^{\ell+1}$, respectively.
The odd and even sectors are also referred to as ``axial'' and ``polar'' and we shall use the two notations indistinctly.
The functions $(H_0,H_1,H_2,K,G,\eta_0,\eta_1)^{\ell}$ and $(h_0,h_1,h_2)^{\ell}$ only depend on $t$ and $r$ and describe the polar parity metric
perturbations and the axial parity metric perturbations, respectively. 
In the following, we adopt the Regge-Wheeler gauge~\cite{Regge:1957td} and set $\eta_i^\ell\equiv G^\ell\equiv h_2^\ell\equiv0$ through a gauge choice.

Using this decomposition, we can solve vacuum Einstein's equations perturbatively in the spin. Because of the transformation properties of the perturbation functions, the linearized equations naturally separate into three groups~\cite{Kojima:1992ie,Pani:2013pma}. 
By denoting the linearized Einstein equations as $\delta {\cal E}_{\mu\nu}=0$, the first group reads
\begin{equation}
\delta{\cal E}_{(I)}\equiv (A^{(I)}_{{\ell}}+{\tilde A}^{(I)}_{{\ell}}\cos\th+\hat{A}^{(I)}_\ell\cos^2\th)Y^{{\ell}}
+(B^{(I)}_{{\ell}}+\tilde B^{(I)}_{{\ell}}\cos\th)\sin\th Y^{{\ell}}_{,\vartheta}=0,\label{eqG1}
\end{equation}
where $I=0,1,2,3$ corresponds to $\delta {\cal E}_{tt}=0$, $\delta
{\cal E}_{tr}=0$, $\delta {\cal E}_{rr}=0$ and $\delta {\cal
  E}_{\vartheta\vartheta} +{\delta {\cal
    E}_{\varphi\varphi}}/{\sin\vartheta^2}=0$, respectively.
The second group reads
\begin{eqnarray}
&\delta{\cal E}_{(L\vartheta)}&\equiv(\alpha^{(L)}_{{\ell}}+{\tilde \alpha}^{(L)}_{{\ell}}\cos\th+{\hat\alpha}^{(L)}_\ell\cos^2\th) Y^{{\ell}}_{,\vartheta}-
(\beta^{(L)}_{{\ell}}+{\tilde \beta}^{(L)}_{{\ell}}\cos\th+\hat{\beta}^{(L)}_\ell\cos^2\th)\frac{ Y^{{\ell}}_{,\varphi}}{\sin\th}\nn\\
&&+(\eta^{(L)}_{{\ell}}+\tilde\eta^{(L)}_\ell\cos\th)\sin\th Y^{{\ell}}+(\xi^{(L)}_{{\ell}}+\tilde\xi^{(L)}_{{\ell}}\cos\th)X^{{\ell}}+
\chi^{(L)}_{{\ell}}\sin\th W^{{\ell}}=0,\label{eqG2a}\\
&\delta{\cal E}_{(L\varphi)}&\equiv(\beta^{(L)}_{{\ell}}+{\tilde \beta}^{(L)}_{{\ell}}\cos\th+\hat{\beta}^{(L)}_\ell\cos^2\th+\tilde\Delta^{(L)}_\ell\sin^2\th) Y^{{\ell}}_{,\vartheta}+
(\alpha^{(L)}_{{\ell}}+{\tilde \alpha}^{(L)}_{{\ell}}\cos\th+{\hat\alpha}^{(L)}_\ell\cos^2\th+\Delta^{(L)}_\ell\sin^2\th)\frac{ Y^{{\ell}}_{,\varphi}}{\sin\th}\nn\\
&&+(\zeta^{(L)}_{{\ell}}+\tilde\zeta^{(L)}_{{\ell}}\cos\th)\sin\th Y^{{\ell}}+\chi^{(L)}_{{\ell}}X^{{\ell}}-
(\xi^{(L)}_{{\ell}}+\tilde\xi^{(L)}_{{\ell}}\cos\th)\sin\th W^{{\ell}}=0,\label{eqG2b}
\end{eqnarray}
where $L=0,1$ and the first equation corresponds to $\delta {\cal
  E}_{t\vartheta}=0$ and $\delta {\cal E}_{r\vartheta}=0$, whereas the
last equation corresponds to $\delta {\cal E}_{t\varphi}=0$ and
$\delta {\cal E}_{r\varphi}=0$.
Finally the third group is
\begin{eqnarray}
\delta {\cal E}_{(\vartheta\varphi)}&\equiv&	
(f_{{\ell}}+\tilde f_{{\ell}}\cos\th) \sin\th Y^{{\ell}}_{,\th}+(g_{{\ell}}+\tilde g_{{\ell}}\cos\th ) Y^{{\ell}}_{,\varphi}
+k_\ell\sin^2\th Y^\ell
+(s_{{\ell}}+\hat{s}_\ell \cos^2\th)\frac{X^{{\ell}}}{\sin\th}+(t_{{\ell}} +\hat{t}_\ell\cos^2\th) W^{{\ell}}=0\,,\nn\\ \label{eqG3a}\\
\delta {\cal E}_{(-)}&\equiv&
(g_{{\ell}}+\tilde g_{{\ell}}\cos\th) \sin\th Y^{{\ell}}_{,\th}-(f_{{\ell}}+\tilde f_{{\ell}}\cos\th ) Y^{{\ell}}_{,\varphi}
+\tilde{k}_\ell\sin^2\th Y^\ell
-(t_{{\ell}}+\hat{t}_\ell \cos^2\th)\frac{X^{{\ell}}}{\sin\th}+(s_{{\ell}} +\hat{s}_\ell\cos^2\th) W^{{\ell}}=0\,,\nn\\ \label{eqG3b}
\end{eqnarray}
corresponding to $\delta {\cal E}_{\vartheta\varphi}=0$ and $\delta {\cal  E}_{\vartheta\vartheta} -{\delta {\cal E}_{\varphi\varphi}}/{\sin\vartheta^2}=0$, respectively. In the equations above, $X^\ell$ and $W^\ell$ are the tensor spherical harmonics defined as in Eq.~\eqref{XW}.
The coefficients appearing in these equations are \emph{linear} and \emph{purely radial} functions of the perturbation variables. Furthermore, they naturally divide into two sets accordingly to their parity:
 \begin{eqnarray}
 \text{Polar:}\qquad && A^{(I)}_{{\ell}},\quad \hat{A}^{(I)}_{{\ell}},\quad \tilde B^{(I)}_{{\ell}}, \quad \alpha^{(L)}_{{\ell}},\quad \hat\alpha^{(L)}_{{\ell}},\quad \tilde\beta^{(L)}_{{\ell}},\quad \tilde\eta^{(L)}_{{\ell}},\quad \zeta^{(L)}_{{\ell}},\quad \xi^{(L)}_{{\ell}},\quad \Delta^{(L)}_\ell, \quad  f_{{\ell}},\quad \tilde{g}_\ell,\quad s_{{\ell}},\quad \hat s_{{\ell}}, \quad \tilde{k}_\ell \nn\\
 \text{Axial:}\qquad &&\tilde A^{(I)}_{{\ell}},\quad B^{(I)}_{{\ell}},\quad \beta^{(L)}_{{\ell}},\quad \hat\beta^{(L)}_{{\ell}},\quad \tilde\alpha^{(L)}_{{\ell}},\quad \eta^{(L)}_{{\ell}},\quad \tilde\xi^{(L)}_{{\ell}},\quad \chi^{(L)}_{{\ell}},\quad \tilde\zeta^{(L)}_{{\ell}},\quad \tilde\Delta^{(L)}_\ell, \quad  g_{{\ell}},\quad \tilde{f}_\ell,\quad  t_{{\ell}},\quad \hat{t}_\ell, \quad {k}_\ell.\nn
\end{eqnarray}
The explicit form of the
coefficients is given in the online notebook in the Supplemental Material. Note that the coefficients above are purely radial functions, i.e., the entire angular dependence has been completely factored out in the linearized Einstein equations.

\subsubsection{Separation of the angular dependence}
The decoupling of the angular dependence of the Einstein equations for
a slowly-rotating star to first order in the spin was performed by Kojima~\cite{Kojima:1992ie} (see also~\cite{ChandraFerrari91}) by using
the orthogonality properties of the spherical harmonics. The procedure has been extended to the case of slowly-rotating BHs to second order in Refs.~\cite{Pani:2012vp,Pani:2012bp}. Here we
adopt the same technique. 

The decoupling is achieved by computing the following integrals
\begin{eqnarray}
 0&=&\int d\Omega {Y^*}^\ell \delta{\cal E}_{(I)}\,, \\
 0&=&\int d\Omega\left[ {Y^*}^{\ell'}_{,\vartheta}\delta{\cal E}_{(L\vartheta)}+\frac{{Y^*}^{\ell'}_{,\varphi}}{\sin\vartheta}\delta{\cal E}_{(L\varphi)}\right] \,,\\
 0&=&\int d\Omega\left[ {Y^*}^{\ell'}_{,\vartheta}\delta{\cal E}_{(L\varphi)}-\frac{{Y^*}^{\ell'}_{,\varphi}}{\sin\vartheta}\delta{\cal E}_{(L\vartheta)}\right] \,,\\
 0&=&\int d\Omega \frac{1}{{\ell(\ell+1)}-2}\left[{W^*}^{\ell'}\delta{\cal E}_{(-)}+\frac{{X^*}^{\ell'}}{\sin\vartheta}\delta{\cal E}_{(\vartheta\varphi)}\right]\,, \label{int_tensor1}\\
 0&=&\int d\Omega \frac{1}{{\ell(\ell+1)}-2}\left[{W^*}^{\ell'}\delta{\cal E}_{(\vartheta\varphi)}-\frac{{X^*}^{\ell'}}{\sin\vartheta}\delta{\cal E}_{(-)}\right]\,, \label{int_tensor2}
\end{eqnarray}
for $I=0,1,2,3$ and $L=0,1$.
These integrals might be evaluated analytically for generic $(\ell,m)$ by using the properties of the spherical harmonics~\cite{Kojima:1992ie,Pani:2012vp,Pani:2012bp}. The resulting equations have the form~\eqref{eq_axial}--\eqref{eq_polar}. For simplicity, here we perform the integrals above explicitly. For example, for $\ell=m=2$ one obtains the system of equations
\begin{eqnarray}
A_2^{(I)}+\frac{\tilde{A}^{(I)}_3}{\sqrt{7}}+\frac{\hat{A}^{(I)}_2}{7}+\frac{2 \hat{A}^{(I)}_4}{7 \sqrt{3}}-\frac{4 B_3^{(I)}}{\sqrt{7}}+\frac{2 \tilde{B}^{(I)}_2}{7}-\frac{10 \tilde{B}^{(I)}_4}{7 \sqrt{3}}&=&0\,,\\
6 \alpha _2^{(L)}+\frac{8 {\tilde\alpha}^{(L)}_3}{\sqrt{7}}+\frac{10 {\hat\alpha}^{(L)}_2}{7}+\frac{20 {\hat\alpha}^{(L)}_4}{7 \sqrt{3}}-\frac{4 i {\hat\beta}^{(L)}_3}{\sqrt{7}}-2 i {\tilde\beta}^{(L)}_2+4 \Delta _2^{(L)}+\frac{8 i {\tilde\Delta}^{(L)}_3}{\sqrt{7}}-2 i \zeta _2^{(L)}-\frac{2 i {\tilde\zeta}^{(L)}_3}{\sqrt{7}}&&\nn\\
+\frac{2 \eta _3^{(L)}}{\sqrt{7}}+\frac{2 {\tilde\eta}^{(L)}_2}{7}+\frac{4 {\tilde\eta}^{(L)}_4}{7 \sqrt{3}}+8 i \xi_2^{(L)}-\frac{40 \chi _3^{(L)}}{\sqrt{7}} &=&0\,,\\
6 \beta^{(L)} _2 +\frac{8 \tilde{\beta}^{(L)}_3}{\sqrt{7}} +\frac{10 \hat\beta^{(L)}_2}{7} +\frac{20 \hat\beta^{(L)}_4}{7 \sqrt{3}} +\frac{4 i \hat{\alpha}^{(L)}_3}{\sqrt{7}}+2 i \tilde{\alpha}^{(L)}_2+\frac{4 i \Delta^{(L)} _3}{\sqrt{7}}+\frac{4 \tilde{\Delta}^{(L)}_2}{7}-\frac{20 \tilde{\Delta}^{(L)}_4}{7 \sqrt{3}}+\frac{2 \zeta^{(L)} _3}{\sqrt{7}}&&\nn\\
+\frac{2 \tilde{\zeta}^{(L)}_2}{7}+\frac{4 \tilde{\zeta}^{(L)}_4}{7 \sqrt{3}}+2 i \eta^{(L)} _2+\frac{2 i \tilde{\eta}^{(L)}_3}{\sqrt{7}}+\frac{40 \xi^{(L)} _3}{\sqrt{7}}+\frac{40 \sqrt{3} \tilde{\xi}^{(L)}_4}{7}-\frac{24 \tilde{\xi}^{(L)}_2}{7}+8 i \chi^{(L)}_2 &=&0\,,\\
6 s_2-2 i f_2-\frac{3 i {\tilde f}_3}{\sqrt{7}}+\frac{2 g_3}{\sqrt{7}}+\frac{6 {\tilde g}_2}{7}+\frac{5 {\tilde g}_4}{7 \sqrt{3}}-\frac{i k_3}{\sqrt{7}}+\frac{4 {\tilde k}_2}{7}+\frac{{\tilde k}_4}{7 \sqrt{3}}+\frac{22 {\hat s}_2}{7}+\frac{10 \sqrt{3} {\hat s}_4}{7}-\frac{10 i {\hat t}_3}{\sqrt{7}} &=&0\,,\\
6 t_2+\frac{2 f_3}{\sqrt{7}}+\frac{6 {\tilde f}_2}{7}+\frac{5 {\tilde f}_4}{7 \sqrt{3}}+2 i g_2+\frac{3 i {\tilde g}_3}{\sqrt{7}}+\frac{4 k_2}{7}+\frac{k_4}{7 \sqrt{3}}+\frac{i {\tilde k}_3}{\sqrt{7}}+\frac{10 i {\hat s}_3}{\sqrt{7}}+\frac{22 {\hat t}_2}{7}+\frac{10 \sqrt{3} {\hat t}_4}{7} &=&0\,,
\end{eqnarray}
which describe the full set of $\ell=m=2$ perturbations to second order in the spin. Similar equations can be found for other values of $\ell$ and $|m|\leq\ell$.

To summarize, our decoupling procedure in the slow-rotation limit allows us to obtain a system of $10$ coupled, ordinary differential equations. 
The mixing of different angular functions in Eqs.~\eqref{eqG1}-\eqref{eqG3b} yields a mixing of perturbation functions with multipolar indices $\ell$, $\ell+1$ and $\ell-1$ to first order in the spin and with multipolar indices $\ell$, $\ell+2$ and $\ell-2$ to second order, respectively. The explicit form of the final radial equations is available online in the Supplemental Material. 

\subsection{Perturbations with $\ell=1$ and $\ell=0$}
In the derivation presented in this Appendix we have always assumed $\ell\geq2$. However, in the rotating case axial and polar perturbations with $\ell=2$ are respectively coupled to polar and axial perturbations with $\ell=3$ and $\ell=1$ and with axial and polar perturbations with $\ell=0$ and $\ell=4$. Perturbations with $\ell=0,1$ satisfy a different set of equations than the one presented above. The perturbation equations for $\ell=1$ can be found by neglecting Eqs.~\eqref{int_tensor1} and \eqref{int_tensor2} (which are clearly ill-defined when $\ell=1$) and by using a residual gauge freedom in order to set $K^{(1)}=0$ and $h_1^{(1)}=0$ in the ansatz for the metric~\cite{1970ApJ...159..847C,1989ApJ...345..925L}. For $\ell=0$, only scalar polar perturbations exist and one can set $K^{(0)}=0$ and $H_1^{(0)}=0$ without loss of generality.

\section{Sources for the inhomogeneous equations governing tidal deformations}\label{app:sources}
The source terms entering the axial perturbations for $L=3$ and $L=1$ read
\begin{eqnarray}
 S_{A}^{(1)} &=&  -\frac{4 M^2 \chi}{\sqrt{15} r^3 (r-2 M)}  \left(r (r-2 M) {H_0^{(2)}}'+(2 M+3 r) H_0^{(2)}\right)  \,,\\
 S_{A}^{(3)} &=&  \frac{2 M^2 \chi}{\sqrt{35} r^3 (r-2 M)^2}  \left(r \left(18 M^2+2 r^2-13 M r\right) {H_0^{(2)}}'+2 \left(M^2-2 r^2+M r\right) H_0^{(2)}\right)  \,,
\end{eqnarray}
whereas the source term entering Eq.~\eqref{eqBHP2} reads
\begin{eqnarray}
 S_{P}^{(2)} &=& -\frac{2 \chi}{105 M^2 r^7 (r-2 M)^3}  \left[r (r-2 M) \left(5 \chi  {H_0^{(2)}}' \left(9 {\delta q} r^6 (4 r-M) (r-2 M)^2 \tanh ^{-1}\left(\frac{M}{r-M}\right)\right.\right.\right.\nn\\
 &&\left.\left.\left.+M \left(270 M^8-36 {\delta q} r^8+117 {\delta q} M r^7-3 M^2 r^6 (7 {\delta m}+25 {\delta q})-12 {\delta q} M^3 r^5+6 (6-7 {\delta q}) M^4 r^4-42 M^5 r^3-22 M^6 r^2\right.\right.\right.\right.\nn\\
 &&\left.\left.\left.\left. -46 M^7 r\right)\right)-4 \sqrt{5} M^4 r \left(r \left(7 \sqrt{3} \left(15 M^2+r^2-9 M r\right) {h_0^{(1)}}'-9 \sqrt{7} \left(5 M^2+2 r^2-3 M r\right) {h_0^{(3)}}'\right) \right.\right.\right.\nn\\
 &&\left.\left.\left.+14 \sqrt{3} \left(-15 M^2+2 r^2+8 M r\right) {h_0^{(1)}}+9 \sqrt{7} \left(10 M^2+12 r^2-17 M r\right) {h_0^{(3)}}\right)\right)\right.\nn\\
 &&\left.+10 \chi  {H_0^{(2)}} \left(9 {\delta q} r^6 (r-2 M)^2 \left(-10 M^2+3 r^2+6 M r\right) \tanh ^{-1}\left(\frac{M}{r-M}\right)\right.\right. \nn\\
 &&\left.\left.+M \left(-270 M^9-27 {\delta q} r^9+27 {\delta q} M r^8+9 M^2 r^7 (24 {\delta q}-7 {\delta m})+15 M^3 r^6 (7 {\delta m}-24 {\delta q})+6 (14 {\delta q}-3) M^4 r^5\right.\right.\right. \nn\\
 &&\left.\left.\left.+12 (2 {\delta q}+3) M^5 r^4+320 M^6 r^3-678 M^7 r^2+376 M^8 r\right)\right)\right] \,.
\end{eqnarray}
The source term entering Eq.~\eqref{eqBHP4} reads
\begin{eqnarray}
 S_{P}^{(4)} &=& -\frac{4 \chi}{35 M^2 r^7 (r-2 M)^3}  \left[2 r (r-2 M) \left(\sqrt{5} \chi  {H_0^{(2)}}' \left(3 {\delta q} r^6 (M+3 r) (r-2 M)^2 \tanh ^{-1}\left(\frac{M}{r-M}\right)\right.\right.\right.\nn\\
 &&\left.\left.\left.+M \left(15 M^8-9 {\delta q} r^8+24 {\delta q} M r^7-3 {\delta q} M^2 r^6-10 {\delta q} M^3 r^5+(9-14 {\delta q}) M^4 r^4+7 M^5 r^3-135 M^6 r^2+167 M^7 r\right)\right).\right.\right.\nn\\
 &&\left.\left.-5 \sqrt{7} M^4 r^2 \left(5 M^2+2 r^2-3 M r\right) {h_0^{(3)}}'-10 \sqrt{7} M^4 r \left(-5 M^2+8 r^2+5 M r\right) {h_0^{(3)}}\right)\right.\nn\\
 &&\left.-\sqrt{5} \chi  {H_0^{(2)}} \left(18 {\delta q} r^6 (r-2 M)^2 \left(5 M^2+2 r^2-10 M r\right) \tanh ^{-1}\left(\frac{M}{r-M}\right)\right.\right.\nn\\
 &&\left.\left.+M \left(60 M^9-36 {\delta q} r^9+288 {\delta q} M r^8-678 {\delta q} M^2 r^7+486 {\delta q} M^3 r^6-24 M^4 r^5+(4 {\delta q}+223) M^5 r^4\right.\right.\right.\nn\\
 &&\left.\left.\left.-768 M^6 r^3+1112 M^7 r^2-600 M^8 r\right)\right)\right] \,.
\end{eqnarray}
Finally, the source terms entering Eqs.~\eqref{eqP0a}--\eqref{eqP0b} read
\begin{eqnarray}
 S_{P}^{(0,0)} &=& -\frac{\chi}{30 M^2 r^6 (r-2 M)^3}  \left[r (r-2 M) \left(\sqrt{5} \chi  {H_0^{(2)}}' \left(9 {\delta q} r^5 (r-2 M)^2 \left(-6 M^2+6 r^2-M r\right) \tanh ^{-1}\left(\frac{M}{r-M}\right)\right.\right.\right.\nn\\
 &&\left.\left.\left.+M \left(96 M^8-54 {\delta q} r^8+171 {\delta q} M r^7-45 {\delta q} M^2 r^6-186 {\delta q} M^3 r^5+6 (9 {\delta q}-1) M^4 r^4-4 (3 {\delta q}-5) M^5 r^3+118 M^6 r^2\right.\right.\right.\right.\nn\\
 &&\left.\left.\left.\left.-300 M^7 r\right)\right)-20 \sqrt{3} M^4 r^3 (r-2 M) {h_0^{(1)}}'+40 \sqrt{3} M^4 r^2 (3 r-2 M) {h_0^{(1)}}\right)\right.\nn\\
 &&\left.+2 \sqrt{5} \chi  {H_0^{(2)}} \left(9 {\delta q} r^5 (r-2 M)^2 \left(2 M^3+r^3+5 M r^2-7 M^2 r\right) \tanh ^{-1}\left(\frac{M}{r-M}\right)+M \left(-96 M^9-9 {\delta q} r^9-18 {\delta q} M r^8\right.\right.\right.\nn\\
 &&\left.\left.\left.+186 {\delta q} M^2 r^7-273 {\delta q} M^3 r^6+6 (10 {\delta q}+9) M^4 r^5+2 (27 {\delta q}-95) M^5 r^4+6 (2 {\delta q}+21) M^6 r^3+174 M^7 r^2-132 M^8 r\right)\right)\right]\,,\nn\\ \\
 S_{P}^{(0,2)} &=&  \frac{\chi}{10 M^2 r^6 (r-2 M)^2}  \left[-r (r-2 M) \left(\sqrt{5} \chi  {H_0^{(2)}}' \left(3 {\delta q} r^5 \left(12 M^3+6 r^3-5 M r^2-20 M^2 r\right) \tanh ^{-1}\left(\frac{M}{r-M}\right)\right.\right.\right.\nn\\
 &&\left.\left.\left.+M \left(80 M^7-18 {\delta q} r^7-3 {\delta q} M r^6+51 {\delta q} M^2 r^5+8 {\delta q} M^3 r^4+2 ({\delta q}+9) M^4 r^3-64 M^5 r^2+50 M^6 r\right)\right)\right.\right.\nn\\
 &&\left.\left.+20 \sqrt{3} M^4 r^3 {h_0^{(1)}}'-40 \sqrt{3} M^4 r^2 {h_0^{(1)}}\right)\right.\nn\\
 &&\left.-2 \sqrt{5} \chi  {H_0^{(2)}} \left(3 {\delta q} r^5 \left(-4 M^4+9 r^4-21 M r^3-9 M^2 r^2+32 M^3 r\right) \tanh ^{-1}\left(\frac{M}{r-M}\right)\right.\right.\nn\\
 &&\left.\left.+M \left(-80 M^8-27 {\delta q} r^8+36 {\delta q} M r^7+54 {\delta q} M^2 r^6-39 {\delta q} M^3 r^5+6 ({\delta q}-3) M^4 r^4-2 ({\delta q}-51) M^5 r^3\right.\right.\right.\nn\\
 &&\left.\left.\left.-170 M^6 r^2+150 M^7 r\right)\right)\right] \,.
\end{eqnarray}

\section{Explicit solution for a tidally-deformed spinning vacuum geometry to first order in the spin}\label{app:solutions}
Here we present the explicit solution for a tidally-deformed spinning vacuum geometry to first order in the spin. To zeroth order in the spin and in the tidal field, the background solution was given in Appendix~\ref{app:background}. For $L=2$ polar-led tidal perturbations, the nonvanishing metric components to zero order in the spin and to first order in the tidal field are given in Eqs.~\eqref{polL2spin0}--\eqref{K0inf}. Finally, to first order in the spin and in the tidal field, the only nonvanishing metric functions are the following $L=3$ and $L=1$ axial components:
\begin{eqnarray}
 h_0^{(1)}(y)&=&\left[2 \sqrt{15} y^2 \alpha +y^3 {\alpha_1}+{\gamma_1}\right]\frac{ M \chi }{y}\nn\\
 &&+\frac{M \chi}{2 \sqrt{15} y^2}  \left[-4-30 (y-1) y^2-3 y \left(4+5 (y-2) y^2\right) \log\left(1-\frac{2}{y}\right)\right]\,, \\
 h_0^{(3)}(y)&=&  \frac{M \chi }{3360 y^2}\left[192 \sqrt{35} \gamma +y \left(528 \sqrt{35} \gamma -420 \gamma_3+20 \alpha_3 (y-2) (3 y-4) y^3+192 \sqrt{35} \alpha  (5 y-4)\right.\right.\nn\\
 &&\left.\left.-15 (3 (y-2) y (3 y-1)+2) y \left(36 \sqrt{35} \gamma +35 \gamma_3\right)\right) \right]\nn\\
 && -\frac{M \chi }{2240 y}\left[525 \gamma_3 (y-2) (3 y-4) y^3+4 \sqrt{35} \gamma  \left(5 y \left(27 (y-2) y^2 (3 y-4)-16\right)+64\right)\right]\log\left(1-\frac{2}{y}\right)\,,
\end{eqnarray}
where again $y=r/M$ and $\alpha_{1,3}$ and $\gamma_{1,3}$ are integration constants related to the homogeneous problem associated with Eqs.~\eqref{eqAXIALl1} and \eqref{eqAXIALl3} (cf. main text, Eq.~\eqref{decomposition} and Table~\ref{tab:metric}).
An analytical solution can be found in closed form also to second order in the spin, by solving Eqs.~\eqref{eqBHP2}, \eqref{eqBHP4}, \eqref{eqP0a} and \eqref{eqP0b} analytically, but it is too cumbersome to be displayed here. The explicit form is given in a notebook in the Supplemental Material.

The first-order solution generalizes that found by Poisson~\cite{Poisson:2014gka} to the case of generic vacuum and reduces to it in the BH case with $m=0$ and with a purely electric quadrupolar tidal field. To linear order in the spin, the coordinate transformation that brings our BH metric to the axisymmetric, electric-led solution found in~\cite{Poisson:2014gka} reads\footnote{Actually, our solution coincides with that found in~\cite{Poisson:2014gka} after adding a term $\frac{4}{5} \frac{M^4}{r^4}$ in the function $f_4^{\rm d}$ defined in Table III of Ref.~\cite{Poisson:2014gka}. Such term, however, is arbitrary as it can be reabsorbed in a shift of the body's angular-momentum vector. We thank Eric Poisson and Phil Landry for clarifying this point in a private communication.}
\begin{eqnarray}
 t &\to& v-r_* -\alpha\sqrt{\frac{5}{\pi }}\frac{   r^3  [1+3 \cos (2 \vartheta)]}{8 M^2} \,,\label{trasft}\\
 r &\to& r \,,\\
 \vartheta &\to & \vartheta + 3\alpha \sqrt{\frac{5}{\pi }}\frac{ \sin (2 \vartheta) \left(2 M^2-r^2\right)}{8 M^2}\,,\label{trasfr}\label{trasfth}\\
 \varphi &\to& \varphi +\alpha  \chi  \sqrt{\frac{5}{\pi }}\frac{3  v}{2 M} - \chi   \left[\tanh ^{-1}\left(\frac{M}{M-r}\right)+\frac{M}{r}\right] +\frac{3 \alpha  \chi }{4 \sqrt{5 \pi } r}\left\{M-5 M \cos (2 \vartheta)+2r\left[\log\left(\frac{r}{M}\right)-11\log \left(2-\frac{r}{M}\right)\right]\right\}\,,\nn\\ \label{trasfph}
\end{eqnarray}
where $dr/dr_*=1-2M/r$ and the term proportional to the light-cone time $v$ in $\varphi$ is needed to eliminate the gauge term proportional to $\gamma^{\rm d}$ in Ref.~\cite{Poisson:2014gka}. In the notation of Ref.~\cite{Poisson:2014gka}, ${\cal E}_0^q= \frac{3}{4} \sqrt{\frac{5}{\pi }} \frac{\alpha}{M^2}$, $M$ and $\chi$ are the same quantities as the ones we used, and we remark that the normalization of the spherical harmonics in Ref.~\cite{Poisson:2014gka} differs from ours.

\section{Low-velocity expansion of the energy change $\Delta E$}\label{app:DeltaE}
For our tidally-distorted vacuum solution, the low-velocity expansion of the energy change $\Delta E$ per logarithmic interval of the orbital frequency, Eq.~\eqref{DeltaERyan}, reads
\begin{eqnarray}
 &\Delta E &= \left[\frac{M}{3 {M_0}}+\frac{{\delta m} M \chi ^2}{3 {M_0}}-\frac{\left({\gamma_0}+12 \sqrt{5} \gamma  {\delta q}\right) M \chi
   ^2}{12 {M_0} \sqrt{\pi }}\right]v^2+\left[-\frac{M^2}{2 {M_0}^2}-\frac{{\delta m} M^2 \chi ^2}{{M_0}^2}+\frac{\left({\gamma_0}+12 \sqrt{5} \gamma  {\delta q}\right)
   M^2 \chi ^2}{4 {M_0}^2 \sqrt{\pi }}\right]v^4\nn\\
   &&+\left[\frac{2 \left(9 {M_0}^{1/6}+M^{1/6}\right) M^{5/2} \chi }{9 {M_0}^{8/3}}+\frac{\left(9 {M_0}^{1/6}+M^{1/6}\right) M^{5/2} \left(-2 \sqrt{5
   \pi } \gamma +\sqrt{3 \pi } {\gamma_1} M^2\right) \chi }{18 {M_0}^{8/3} \pi }\right]v^5\nn\\
   &&+\left[-\frac{27 M^3}{8 {M_0}^3}-\frac{\gamma  \left(5 {M_0}^{1/6}+M^{1/6}\right) M^3}{3 {M_0}^{19/6} \sqrt{5 \pi }}-\frac{\left(5 {M_0}^{1/6}
   (20+243 {\delta m}-16 {\delta q})+4 (5-4 {\delta q}) M^{1/6}\right) M^3 \chi ^2}{120 {M_0}^{19/6}}\right.\nn\\
   &&\left.-\frac{M^3 \chi ^2}{10080 {M_0}^{19/6} \sqrt{\pi }}\left(15
   {M_0}^{1/6} \left(-1701 {\gamma_0}+224 \sqrt{5} {\gamma_2}+390 \sqrt{7} {\gamma_3}+4 \sqrt{5} \gamma  (270+112 {\delta m}-4951 {\delta q})\right)\right.\right.\nn\\
   &&\left.\left.+2 \left(336 \sqrt{5} {\gamma_2}+585 \sqrt{7} {\gamma_3}+4 \sqrt{5} \gamma  (405+182 {\delta m}+228 {\delta q})\right) M^{1/6}\right)\right]v^6\nn\\
   &&+\left[\frac{2 \left(15 {M_0}^{1/6}-M^{1/6}\right) M^{7/2} \chi }{3 {M_0}^{11/3}}+\frac{\left(15 {M_0}^{1/6}-M^{1/6}\right) M^{7/2} \left(-2 \sqrt{5
   \pi } \gamma +\sqrt{3 \pi } {\gamma_1} M^2\right) \chi }{6 {M_0}^{11/3} \pi }\right]v^7\nn\\
   &&+\left[-\frac{225 M^4}{16 {M_0}^4}+\frac{4 \gamma  \left(-36 {M_0}^{1/6}+M^{1/6}\right) M^4}{9 {M_0}^{25/6} \sqrt{5 \pi }}+\frac{M^4\chi ^2}{540
   {M_0}^{13/3}} \left(-9
   {M_0}^{1/3} (520+3375 {\delta m}-384 {\delta q})+180 M^{1/3}\right.\right.\nn\\
   &&\left.\left.+4 (475-24 {\delta q}) ({M_0} M)^{1/6}\right) +\frac{M^4 \chi ^2}{15120 {M_0}^{13/3} \sqrt{\pi }} \left(-9 {M_0}^{1/3} \left(-23625 {\gamma_0}+5376 \sqrt{5} {\gamma_2}+9360 \sqrt{7} {\gamma_3}\right.\right.\right.\nn\\
   &&\left.\left.\left.+4
   \sqrt{5} \gamma  (6200+4032 {\delta m}-67227 {\delta q})+560 \sqrt{3} {\gamma_1} M^2\right)+4 \left(3 \left(112 \sqrt{5}
   {\gamma_2}+195 \sqrt{7} {\gamma_3}\right) ({M_0} M)^{1/6}\right.\right.\right.\nn\\
   &&\left.\left.\left.+4 \sqrt{5} \gamma  \left(-315 M^{1/3}+2 (-1355+133 {\delta m}+114
   {\delta q}) ({M_0} M)^{1/6}\right)+70 \sqrt{3} {\gamma_1} \left(9 M^{7/3}+89 \left({M_0}
   M^{13}\right)^{1/6}\right)\right)\right) \right] v^8+\nn\\
   &&\left[\frac{27 \left(7 {M_0}^{1/6}-M^{1/6}\right) M^{9/2} \chi }{4 {M_0}^{14/3}}+\frac{M^{9/2}\chi
   }{5040 {M_0}^{31/6} \sqrt{\pi }} \left(-1120 \sqrt{5} \gamma  ({M_0}
   M)^{1/3}\right.\right.\nn\\
   &&\left.\left.+\sqrt{{M_0}} M^{1/6} \left(11026 \sqrt{5} \gamma +180 \sqrt{7} {\gamma_3}-8505 \sqrt{3} {\gamma_1}
   M^2\right)+{M_0}^{2/3} \left(-76542 \sqrt{5} \gamma +3060 \sqrt{7} {\gamma_3}+59535 \sqrt{3} {\gamma_1} M^2\right)\right) \right]v^9 +\nn\\
   &&\left[-\frac{6615 M^5}{128 {M_0}^5}+\frac{\gamma  \left(-14871 {M_0}^{1/6}+1781 M^{1/6}\right) M^5}{168 {M_0}^{31/6} \sqrt{5 \pi }}-\frac{M^5\chi ^2}{40320 {M_0}^{16/3}}
   \left(3 {M_0}^{1/3} (820120+3472875 {\delta m}-475872 {\delta q})\right.\right.\nn\\
   &&\left.\left.+8 \left(6160 M^{1/3}+(-149555+21372 {\delta q}) ({M_0}
   M)^{1/6}\right)\right)-\frac{M^5 \chi ^2}{806400 {M_0}^{16/3} \sqrt{\pi }} \left({M_0}^{1/3} \left(4 \sqrt{5} \gamma  (13449460+14276160 {\delta m}\right.\right.\right.\nn\\
   &&\left.\left.\left.-146740863 {\delta q})+5 \left(-10418625 {\gamma_0}+2855232 \sqrt{5} {\gamma_2}+4980820 \sqrt{7} {\gamma_3}-249600
   {\gamma_4}+1351680 \sqrt{3} {\gamma_1} M^2\right)\right)\right.\right.\nn\\
   &&\left.\left.-4 \left(4 \sqrt{5} \gamma  \left(-616 (-115+12 {\delta q}) M^{1/3}+5
   (-131085+89050 {\delta m}+50508 {\delta q}) ({M_0} M)^{1/6}\right)\right.\right.\right.\nn\\
   &&\left.\left.\left.+5 \left(\left(85488 \sqrt{5} {\gamma_2}+148655 \sqrt{7}
   {\gamma_3}+4800 {\gamma_4}\right) ({M_0} M)^{1/6}+160 \sqrt{3} {\gamma_1} \left(-154 M^{7/3}+3071 \left({M_0}
   M^{13}\right)^{1/6}\right)\right)\right)\right)\right]v^{10} \nn\\
   &&+{\cal O}(v^{11})\,. \label{DeltaE}
\end{eqnarray}
By comparing Eq.~\eqref{DeltaE} with Eq.~\eqref{DeltaERyan}, we can identify the first multipole moments, Eqs.~\eqref{moment0}--\eqref{moment4}. Although not displayed here, we have also computed the coefficients of $v^{11}$ and $v^{12}$ and checked that they agree with the expansion~\eqref{DeltaERyan} once the definitions~\eqref{moment0}--\eqref{moment4} are imposed. This is a consistency check of our solution.

\section{Explicit solution for a tidally-deformed spinning BH to second order in the spin}\label{app:BHsol}
When the central object is a BH, regularity at the horizon simplifies the metric perturbations considerably and the full solution can be presented in compact form. The nonvanishing components of the line element~\eqref{metric} in the case of a spinning BH deformed by a stationary, axisymmetric quadrupole-led tidal field to second order in the spin read
\begin{eqnarray}
 g_{tt} &=&  -1+\frac{2}{y}+\frac{\chi ^2}{y^5} \left(4-2 y^2+\left(-6+y+3 y^2\right) \sin^2\vartheta\right)+\frac{3}{8} \sqrt{\frac{5}{\pi }} (y-2)^2 \alpha  [1+3 \cos(2\vartheta)]\nn\\
 &&+\frac{3 \alpha  \chi ^2}{64 \sqrt{5 \pi } y^5} \left[104+112 y+344
   y^2-773 y^3+558 y^4+4 \left(-104+104 y+152 y^2-305 y^3+30 y^4\right) \cos(2\vartheta)\right.\nn\\
   &&\left.+5 \left(-40+48 y+40 y^2-75 y^3+18 y^4\right]
   \cos(4\vartheta)\right)\,,\\
 g_{t\varphi} &=& -\frac{2 M \chi  \sin\vartheta^2}{y}-\frac{3 \alpha  M \chi}{4 \sqrt{5 \pi } y}  \left(-12+15 y+20 y^2+5 (-4+5 y) \cos(2\vartheta)\right)
   \sin\vartheta^2\,,\\
 g_{rr} &=& \frac{y}{y-2}-\frac{\chi ^2 \left(10-3 y+y^2+3 \left(10-7
   y+y^2\right) \cos(2\vartheta)\right)}{2 (y-2)^2 y^3}+\frac{3}{8} \sqrt{\frac{5}{\pi }} y^2 \alpha  [1+3 \cos(2\vartheta)]\nn\\
   &&+\frac{3 \alpha  \chi ^2}{448 \sqrt{5 \pi } (y-2)^2 y^3} \left[-896 \left(10-15 y+7 y^2\right)+32 (y-2)
   \left(-55-4 y+20 y^2+10 y^3\right) [1+3 \cos(2\vartheta)]\right.\nn\\
   &&\left.-(y-2) \left(100+62 y-93 y^2+6 y^3\right) (9+20 \cos(2\vartheta)+35
   \cos(4\vartheta))\right]\,,\\
 g_{\vartheta\vartheta} &=& y^2 M^2-\frac{(2+y) M^2 \chi ^2 (1+3
   \cos(2\vartheta))}{2 y^2}+\frac{3}{8} \sqrt{\frac{5}{\pi }} y^2 (y^2-2) \alpha  M^2 [1+3 \cos(2\vartheta)]\nn\\
   &&-\frac{\alpha  M^2 \chi ^2}{64 \sqrt{5 \pi } y^2} \left[-1008+36 y+804 y^2+321 y^3+4 \left(-672+44 y+540 y^2+225 y^3\right)
   \cos(2\vartheta)\right.\nn\\
   &&\left.+5 \left(-336-68 y+252 y^2+63 y^3\right) \cos(4\vartheta)\right]  \,, \label{BHgthethatheta}\\
 g_{\varphi\varphi} &=& g_{\vartheta\vartheta} \sin^2\vartheta  \,,
\end{eqnarray}
where $y=r/M$. This solution generalizes that found by Poisson~\cite{Poisson:2014gka} to quadratic order in the spin, but restricting to the axisymmetric ($m=0$) case with a purely electric quadrupolar tidal field. The coordinate transformation between our solution at that found in Ref.~\cite{Poisson:2014gka} is given in Appendix~\ref{app:solutions}.

\end{widetext}
\bibliography{tidalrot}
\end{document}